\documentclass[11pt]{article}
\usepackage{epsfig,dsfont}
\usepackage{geometry}
\usepackage[T1]{fontenc}
\usepackage[utf8]{inputenc}
\usepackage[english]{babel}
\usepackage{amsmath}
\usepackage{amsfonts}  
\usepackage{epsfig}
\usepackage{calc}
\usepackage{graphicx}
\usepackage{xcolor} 
\usepackage{cite}

\DeclareMathOperator{\Tr}{Tr}

\begin{document}  
\sffamily

\vspace*{1mm}

\begin{center}

{\LARGE
Abelian color cycles: a new approach to strong coupling expansion 
\\
\vskip2mm
and dual representations for non-abelian lattice gauge theory}
\vskip10mm
Christof Gattringer and Carlotta Marchis 
\vskip8mm
Universit\"at Graz, Institut f\"ur Physik, Universit\"atsplatz 5, 8010 Graz, Austria
\end{center}
\vskip15mm

\begin{abstract}
We propose a new approach to strong coupling series and dual representations for 
non-abelian lattice gauge theories using the SU(2) case as an example. 
The Wilson gauge action is written as a sum over "abelian color cycles" (ACC) which 
correspond to loops in color space around plaquettes. The ACCs are complex numbers 
which can be commuted freely such that the strong coupling series and the dual representation 
can be obtained as in the abelian case. Using a suitable representation of the SU(2) gauge
variables we integrate out all original gauge links and identify the 
constraints for the dual variables in the SU(2) case. We show that the construction can be
generalized to the case of SU(2) gauge fields with staggered fermions. The result is a 
strong coupling series where all gauge integrals are known in closed form and we discuss
its applicability for possible dual simulations. The abelian color cycle concept can be generalized
to other non-abelian gauge groups such as SU(3).
\end{abstract}

\vskip10mm

\section{Introduction}

Reformulating physical theories with different variables is an important tool for understanding quantum field theories.
Usually it is not clear which variables are suited best for describing various physical phenomena and often 
different questions require different representations. In particular when using numerical simulations in the framework of
lattice field theories, novel representations may give rise to new simulation strategies which allow one to 
explore parameter regions that were not accessible before (see, e.g., the reviews on worldline and dual 
representations in lattice field theories \cite{Chandrasekharan:2008gp,deForcrand:2010ys,Gattringer:2014nxa}). 

A class of models where alternative representations were studied in various forms are non-abelian lattice gauge theories. 
Despite many interesting ideas, so far we have not seen a major breakthrough concerning computational or conceptual
aspects. Previous work has mainly used strong coupling expansion techniques based on the character expansion,
often combined with the introduction of dual variables \cite{Drouffe:1978dn,Savit:1979ny}. 
Examples for deriving such dual representations and the discussion of their properties are, e.g., given in 
\cite{Anishetty:1990en,Anishetty:1992xa,Halliday:1994he,Oeckl:2000hs,Cherrington:2007ax,Cherrington:2009ak,Cherrington:2009am},
and various strategies for corresponding numerical simulations are explored in 
\cite{HariDass:1999kx,HariDass:2000ca,HariDass:2000tp,Cherrington:2007is,Cherrington:2008ey,
Cherrington:2008mf,Cherrington:2009fx}. Results from directly simulating the leading terms of the 
strong coupling expansion of lattice QCD were presented in 
\cite{deForcrand:2009dh,Fromm:2011kq,deForcrand:2013ufa,deForcrand:2014tha,deForcrand:2015daa} (see also 
\cite{Vairinhos:2014uxa} for another strategy based on Hubbard-Stratonovich transformations).

In this paper we present a new approach for representing non-abelian lattice field theories in terms of dual variables, using
the example of SU(2) lattice gauge theory.  As in previous work the approach is based on a strong coupling expansion, 
but instead of using character expansion of the 
Boltzmann factor, we decompose the gauge action into abelian terms, which have the interpretation of 
loops in color space around plaquettes. We refer to these loops as abelian color cycles (ACC). Using this decomposition, one
can proceed with the dualization as in the abelian case since all involved terms commute. The original gauge fields
are integrated out and the corresponding Haar measure integrals can all be solved in closed form. The invariance
under gauge transformations in the original formulation is converted into constraints for the new dual variables, 
which are integer valued occupation numbers for the ACCs. We show that the representation can be generalized to 
the case of gauge fields coupled to fermions and derive the dual representation which is the exact rewriting of the 
partition sum for SU(2) lattice gauge fields coupled to staggered fermions. We furthermore discuss the representation 
of observables in the dual form and compute the leading terms of a coupled strong coupling and hopping expansion.

\section{SU$(2)$ lattice gauge theory}

The model we use for developing the ACC approach is SU(2) lattice gauge theory with Wilson action
\begin{equation}
S_G[U] \; = \; -\dfrac{\beta}{2} \sum_{x,\mu < \nu} 
\Tr U_{x,\mu} \; U_{x+\hat{\mu},\nu} \, U_{x+\hat{\nu},\mu}^{\dagger} \, U_{x,\nu}^\dagger \; ,
\label{eq:action}
\end{equation}
where the link variables $U_{x,\mu} \in $ SU(2) are the dynamical degrees of freedom which live on the links 
$(x,\mu)$ of a $4$-dimensional lattice with periodic boundary conditions, i.e., a 4-torus 
(the generalization to other dimensions is trivial). The partition sum 
$Z = \int \! D[U] \, e^{-S_G[U]}$ is 
obtained by integrating the Boltzmann factor $e^{-S_G[U]}$ with the product of the invariant Haar measures 
$\int \! D[U] = \prod_{x,\mu} \int_{\text{SU}(2)} dU_{x,\mu}$.

The key step for our new approach is to write the trace and the matrix multiplications in 
(\ref{eq:action}) as explicit sum \footnote{A similar strategy of writing explicitly the summation over internal indices in the action for the CP(N-1) model, combined with introducing individual dual variables for all index combinations, was used in \cite{Chandrasekharan:2008gp}.}. This decomposes the action in the form,  
\begin{equation}
\label{eq:abelianaction}
S_G[U] \; = \; -\dfrac{\beta}{2} \sum_{x,\mu < \nu} \sum_{a,b,c,d=1}^{2} 
U_{x,\mu}^{ab} U_{x+\hat{\mu},\nu}^{bc} U_{x+\hat{\nu},\mu}^{dc \ \star} U_{x,\nu}^{ad \ \star}.
\end{equation}
The products 
$U_{x,\mu}^{ab} U_{x+\hat{\mu},\nu}^{bc} U_{x+\hat{\nu},\mu}^{dc \ \star} U_{x,\nu}^{ad \ \star}$ of link matrix 
components $U_{x,\mu}^{ab}$ are now referred to as the {\sl ''Abelian Color Cycles"} (ACC).
The ACCs are complex numbers and correspond to 
a path in color space closing around a plaquette. At each of the four corners of the plaquette the corresponding four 
color indices $(a, b, c, d)$ can be 1 or 2, such that we have a total of $2^4 = 16$ different color cycles labelled by 
the space time coordinates $(x, \mu < \nu)$ of the plaquette and the values of the color indices $(a,b,c,d)$ at the four corners. 

\begin{figure}[t!]
\begin{center}
\vskip5mm
\includegraphics[scale=0.6,clip]{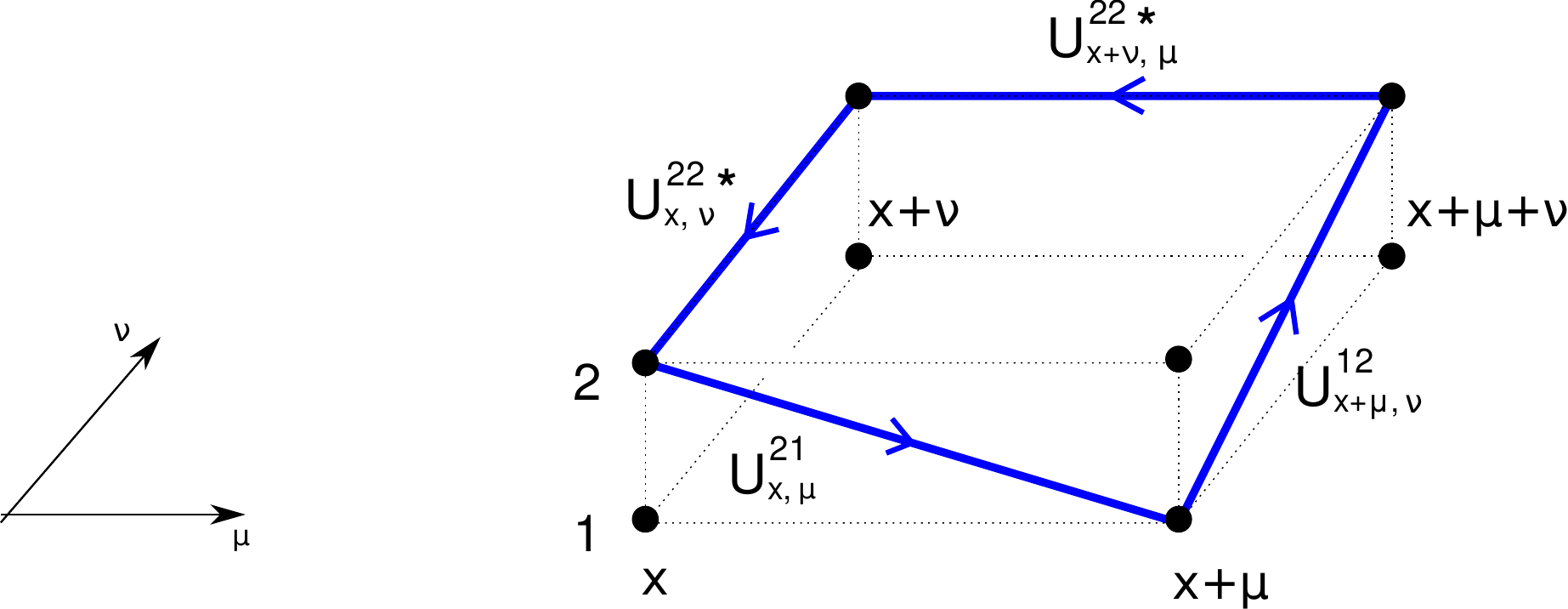}
\end{center}
\caption{The geometrical representation of an abelian color cycle (ACC) in the $\mu$-$\nu$ plane for the example 
$U_{x,\mu}^{21} U_{x+\hat{\mu},\nu}^{12} U_{x+\hat{\nu},\mu}^{22 \ \star} U_{x,\nu}^{22 \ \star}$. Each link element 
$U_{x,\mu}^{ab}$ is represented as an arrow connecting color $a$ at site $x$ to color $b$ at site $x + \hat{\mu}$ and we 
use two layers of the lattice to represent the two colors. The ACC shown here  
corresponds to the cycle occupation number $p_{x,\mu\nu}^{2122} \in \mathds{N}_0$.}
\label{fig:ACC}
\begin{center}
\vskip5mm
\includegraphics[scale=1.23,clip]{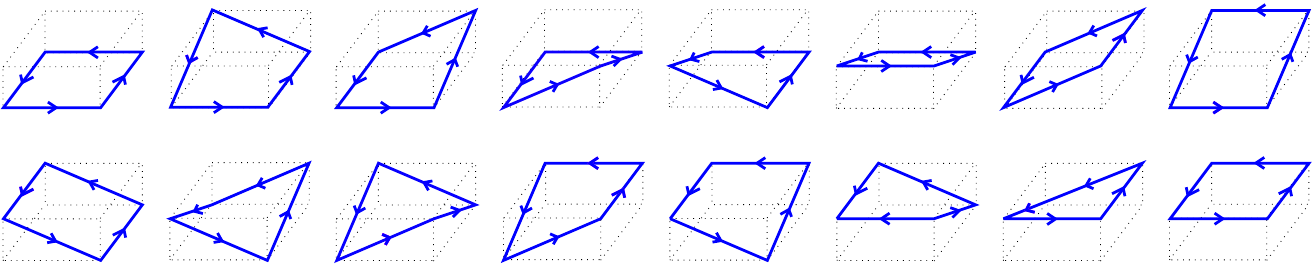}
\end{center}
\caption{The 16 possible abelian color cycles which are attached to a given plaquette. In the dual representation 
their occupation is given by the corresponding cycle occupation number 
$p_{x,\mu\nu}^{abcd} \in \mathds{N}_0$.}
\label{fig:allcycles}	
\end{figure}

It is useful to associate a geometrical representation to the  ACCs, which we illustrate in Fig.~\ref{fig:ACC} for the 
example of the 2122 ACC given by
$U_{x,\mu}^{21} U_{x+\hat{\mu},\nu}^{12} U_{x+\hat{\nu},\mu}^{22 \ \star} U_{x,\nu}^{22 \ \star}$.
The lower left corner of the plaquette is $x$, and at $x$, as well as on the other 3 corners of the plaquette we use
two layers, labelled with 1 and 2 to depict the color at each corner. The first link matrix element  $U_{x,\mu}^{21}$
in the 2122 ACC is represented as an arrow that connects color 2 at $x$ to color 1 at $x + \hat{\mu}$. The next factor is
$U_{x+\hat{\mu},\nu}^{12}$, which consequently is represented by an arrow which connects flavor 1 at $x + \hat{\mu}$
with flavor 2 at 
$x + \hat{\mu} + \hat{\nu}$. The other two factors $U_{x+\hat{\nu},\mu}^{22 \ \star}$ and $U_{x,\nu}^{22 \ \star}$
are complex conjugate, which in our representation corresponds to arrows that run in negative direction. Thus
the two factors $U_{x+\hat{\nu},\mu}^{22 \ \star}$ and $U_{x,\nu}^{22 \ \star}$ close the ACC around the plaquette.
Note that for SU(2), where the trace of the plaquette is already real such that one does not need to
explicitly project to the real part in the action (\ref{eq:action}), only plaquettes with mathematically positive orientation
are needed\footnote{This is different for many other groups, e.g., SU(3), 
where we explicitly have to take the real part in the definition of the action and 
both orientations are needed for the sum of the two complex conjugate terms in the action.}. Thus for SU(2) 
all ACCs are 
oriented in the mathematically positive sense. Using the geometrical representation illustrated in Fig.~\ref{fig:ACC}, 
in Fig.~\ref{fig:allcycles} we show all 16 ACCs possible at a given plaquette.

With the representation (\ref{eq:abelianaction}) of the action as a sum over ACCs, the Boltzmann factor can be 
factorized accordingly and the exponentials of the individual ACCs can be expanded, 
\begin{align}
Z  & =  \int \! D[U] \, e^{-S_G[U]} \; = \; \int \! D[U] \prod_{x,\mu<\nu} \prod_{a,b,c,d = 1}^{2} 
e^{\frac{\beta}{2} U_{x,\mu}^{ab} U_{x+\hat{\mu},\nu}^{bc} U_{x+\hat{\nu},\mu}^{dc \ \star} U_{x,\nu}^{ad \ \star}} 
\nonumber \\
& = \int \! D[U] \prod_{x,\mu<\nu} \prod_{a,b,c,d = 1}^{2} \sum_{p_{x,\mu\nu}^{abcd} = 0}^{\infty} 
\dfrac{ \left( \beta/2 \right)^{p_{x,\mu\nu}^{abcd}}}{p_{x,\mu\nu}^{abcd}\, !} 
\left( U_{x,\mu}^{ab} U_{x+\hat{\mu},\nu}^{bc} U_{x+\hat{\nu},\mu}^{dc \ \star} 
U_{x,\nu}^{ad \ \star} \right)^{p_{x,\mu\nu}^{abcd}} \; .
\label{eq:partitionsum}
\end{align}
In the first step we write the Boltzmann factor as a product over all plaquettes (product over $x,\mu < \nu$) and all 
ACCs on that plaquette (product over $a,b,c,d$). Each individual exponential is then expanded in a power series. 
The corresponding expansion indices $p_{x,\mu\nu}^{abcd} \in \mathds{N}_0$ will turn out to be our dual 
variables. We refer to the  $p_{x,\mu\nu}^{abcd}$ as the {\it ''cycle occupation numbers''}.	
	
Since the ACCs $U_{x,\mu}^{ab} U_{x+\hat{\mu},\nu}^{bc} U_{x+\hat{\nu},\mu}^{dc \ \star} U_{x,\nu}^{ad \ \star}$,
and the link variable matrix elements $U_{x,\mu}^{ab}$ they are made of, are complex numbers, we can commute 
and reorder them. In this way we can determine the integer valued powers $N_{x,\mu}^{ab}$ and 
$\overline{N}_{x,\mu}^{ab}$ for the all matrix elements $U_{x,\mu}^{ab}$ and 
$U_{x,\mu}^{ab \ \star}$. The partition sum turns into
\begin{equation}
Z \; = \; \sum_{\{p\}} \left[ \prod_{x,\mu<\nu} \prod_{a,b,c,d}  
\dfrac{ \left(\beta/2 \right)^{p_{x,\mu\nu}^{abcd}}}{p_{x,\mu\nu}^{abcd}\, !} \right] 
\prod_{x,\mu} \int \! \! dU_{x,\mu} \; \prod_{a,b} \left( U_{x,\mu}^{ab} \right) ^{N_{x,\mu}^{ab}} 
\left( U_{x,\mu}^{ab \ \star} \right) ^{\overline{N}_{x,\mu}^{ab}} \; ,
\label{eq:partitionsum2}
\end{equation}
where we introduced 
$\sum_{\{p\}} = \prod_{x,\mu<\nu} \prod_{a,b,c,d = 1}^{2} \sum_{p_{x,\mu\nu}^{abcd} = 0}^{\infty}$ to
denote the sum over all configurations of the cycle occupation numbers $p_{x,\mu\nu}^{abcd} \in \mathbb{N}_{0}$. 
After the reordering the exponents $N_{x,\mu}^{ab}$ and 
$\overline{N}_{x,\mu}^{ab}$ for the components $U_{x,\mu}^{ab}$ and $U_{x,\mu}^{ab \ \star}$ are obtained as
\begin{equation} 
N_{x,\mu}^{ab} \; = \; 
\sum_{\nu:\mu<\nu}p_{x,\mu\nu}^{abss} + \sum_{\rho:\mu>\rho}p_{x-\hat{\rho},\rho\mu}^{sabs}
\quad , \quad 
\overline{N}_{x,\mu}^{ab} \; = \; 
\sum_{\nu:\mu<\nu}p_{x-\hat{\nu},\mu\nu}^{ssba} + \sum_{\rho:\mu>\rho}p_{x,\rho\mu}^{assb} \; ,
\end{equation} 
where we introduced the label $s$ to indicate independent summation 
over color indices replaced by an $s$. Examples are: 
$p_{x,\mu\nu}^{abss} \equiv \sum_{c,d} p_{x,\mu\nu}^{abcd}$ or 
$p_{x,\mu\nu}^{sabs} \equiv \sum_{c,d} p_{x,\mu\nu}^{cabd}$. 

For obtaining the final form of the partition sum we still have to perform the integration over the Haar measure. 
To do so we choose the following parametrization for the SU(2) gauge links, 
\begin{equation}
\label{eq:parametrization}
U_{x,\mu} =\left(
\begin{array}{cc}
\cos\theta_{x,\mu} \, e^{i\alpha_{x,\mu}}  & \sin\theta_{x,\mu} \, e^{i\beta_{x,\mu}}\\
-\sin\theta_{x,\mu} \, e^{-i\beta_{x,\mu}} & \cos\theta_{x,\mu} \, e^{-i\alpha_{x,\mu}}
\end{array} \right) \; ,
\end{equation}
with $\alpha_{x,\mu} , \beta_{x,\mu} \in [-\pi,\pi]$ and $\theta_{x,\mu} \in [0,\pi/2]$. The normalized Haar measure for this 
representation is easily obtained as 
\begin{equation}
dU_{x,\mu} \; = \; 2 \, d\theta_{x,\mu}\sin\theta_{x,\mu} \cos\theta_{x,\mu} \,
\dfrac{d\alpha_{x,\mu}}{2\pi} \, \dfrac{d\beta_{x,\mu}}{2\pi} \; .
\label{eq:haarmeasure}
\end{equation}
Inserting the parameterization (\ref{eq:parametrization}) and the Haar measure (\ref{eq:haarmeasure}) into 
(\ref{eq:partitionsum2}), the partition sum turns into 
\begin{align}
&Z \; = \; \sum_{\{p\}} \! \left[ \prod_{x,\mu<\nu} \prod_{a,b,c,d} 
 \dfrac{ \left( \beta/2 \right)^{p_{x,\mu\nu}^{abcd}}}{p_{x,\mu\nu}^{abcd}!} \right]
 \prod_{x,\mu} (-1)^{J_{x,\mu}^{21}} \; \,
 2 \! \int_{0}^{\pi/2} \!\!\!\!  d\theta_{x,\mu} \, (\cos\theta_{x,\mu})^{1 + S_{x,\mu}^{11} + S_{x,\mu}^{22}} 
 \; (\sin\theta_{x,\mu})^{1 + S_{x,\mu}^{12} + S_{x,\mu}^{21}} 
 \nonumber
 \\
& \hspace{50mm} \times \; 
\int_{0}^{2\pi} \! \dfrac{d\alpha_{x,\mu}}{2\pi} \; e^{i\alpha_{x,\mu}[J_{x,\mu}^{11}-J_{x,\mu}^{22}]} \; 
\int_{0}^{2\pi} \! \dfrac{d\beta_{x,\mu}}{2\pi} \; e^{i\beta_{x,\mu}[J_{x,\mu}^{12}-J_{x,\mu}^{21}]} 	\; .
\label{eq:partitionsum3}
\end{align} 
For a convenient notation we introduce the integer valued fluxes $J_{x,\mu}^{ab}$ and $S_{x,\mu}^{ab}$,
\begin{equation}
\label{eq:Jfluxes}
J_{x,\mu}^{ab} \; = \; N_{x,\mu}^{ab} - \overline{N}_{x,\mu}^{ab} \; = \; 
\sum_{\nu:\mu<\nu}[\, p_{x,\mu\nu}^{abss} - p_{x-\hat{\nu},\mu\nu}^{ssba} \, ] - 
\sum_{\rho:\mu>\rho}[\, p_{x,\rho\mu}^{assb} - p_{x-\hat{\rho},\rho\mu}^{sabs} \, ] \; ,	
\end{equation}
\begin{equation}
\label{eq:Sfluxes}
S_{x,\mu}^{ab} \; = \; N_{x,\mu}^{ab} + \overline{N}_{x,\mu}^{ab} \; = \; 
\sum_{\nu:\mu<\nu}[\, p_{x,\mu\nu}^{abss} + p_{x-\hat{\nu},\mu\nu}^{ssba}\, ] + 
\sum_{\rho:\mu>\rho}[\, p_{x,\rho\mu}^{assb} + p_{x-\hat{\rho},\rho\mu}^{sabs}\, ] \; .
\end{equation}	
	
It is obvious that in the form (\ref{eq:partitionsum3}) of the partition sum all integrals can be solved in closed form. 
The integrals over $\alpha_{x,\mu}$ and $\beta_{x,\mu}$ in the second line of (\ref{eq:partitionsum3}) give rise
to Kronecker deltas (here denoted as $\delta(n)$) 
which enforce constraints for the fluxes $J_{x,\mu}^{ab}$ at all links $(x,\mu)$:
\begin{equation}
J_{x,\mu}^{11} \, - \, J_{x,\mu}^{22} \; = \; 0 \quad \forall \, x, \mu \qquad \mbox{and} \qquad
J_{x,\mu}^{12} \, - \, J_{x,\mu}^{21} \; = \; 0 \quad \forall \, x, \mu \; .
\label{eq:constraints}
\end{equation}
We will discuss these constraints in more detail below, but for now point out that these constraints for the 
$J_{x,\mu}^{ab}$ fluxes imply that the corresponding combinations of $S_{x,\mu}^{ab}$ fluxes are even, i.e.,
\begin{equation}
S_{x,\mu}^{11} \, + \, S_{x,\mu}^{22} \; = \; \mbox{even} \quad \forall \, x, \mu \qquad \mbox{and} \qquad
S_{x,\mu}^{12} \, + \, S_{x,\mu}^{21} \; = \; \mbox{even} \quad \forall \, x, \mu \; .
\label{eq:Scombinations}
\end{equation}	
The evenness of these combinations is easy to establish: If one inserts the explicit expressions for 
$S_{x,\mu}^{ab}$ into (\ref{eq:Scombinations}) and the explicit expressions for 
$J_{x,\mu}^{ab}$ into (\ref{eq:constraints}) then exactly the same combinations of cycle occupation numbers
$p_{x,\mu\nu}^{abcd}$ appear 
on the left hand sides, but in (\ref{eq:constraints}) some of the terms appear with minus signs. These minus signs on the 
left hand side of (\ref{eq:constraints}) can be converted into plus signs by adding 2-times the terms on both sides of
(\ref{eq:constraints}). This converts the left hand sides of (\ref{eq:constraints}) into the left hand sides of 
(\ref{eq:Scombinations}) and all terms that were added to 0 on the right hand sides of (\ref{eq:constraints})
are multiples of 2. This establishes  (\ref{eq:Scombinations}). In addition 
we point out that the combinations in (\ref{eq:Scombinations}) are non-negative as follows from the definition
(\ref{eq:Sfluxes}) of the $S_{x,\mu}^{ab}$ and the fact that $p_{x,\mu\nu}^{abcd} \in \mathbb{N}_{0}$. 

The combinations (\ref{eq:Scombinations}) are of interest since they appear in the integrals over $\theta_{x,\mu}$ in
the first line of Eq.~(\ref{eq:partitionsum3}). These integrals are a well known representation of the beta function $B$,
\begin{equation}
\label{weightintegrals}
2 \! \int_0^{\pi/2} \!\! d \theta \, ( \cos \theta )^{1+n} \, ( \sin \theta )^{1+m} \; = \; 
B\left( \frac{n}{2} \!+\! 1 \Big| \frac{m}{2} \!+\! 1 \right) \; = \; 
\frac{ \Gamma \left( \frac{n}{2} + 1 \right) \, \Gamma \left( \frac{m}{2} + 1 \right) }{
\Gamma \left( \frac{n+m + 2}{2} + 1 \right) } \; = \; 
\frac{ \frac{n}{2} ! \; \frac{m}{2} ! }{ \left(\frac{n+m + 2}{2}\right) ! } \; ,
\end{equation}
where in the third step we represent the beta function $B$ in terms of gamma functions $\Gamma$
and in the last step, assuming that $n$ and $m$ are even, express these in terms of factorials.
	
Putting things together, we can now write the partition sum in the form
\begin{equation}
Z \; = \; \sum_{\{p\}} W_{\beta}[p] \; W_H[p] \; (-1)^{\sum_{x,\mu}J_{x,\mu}^{21}} \; 
\prod_{x,\mu}  \delta(J_{x,\mu}^{11}-J_{x,\mu}^{22}) \; \delta(J_{x,\mu}^{12}-J_{x,\mu}^{21}) \; .
\label{eq:partitionsum4}
\end{equation}
We have introduced two weight factors, the $\beta$-dependent weight factor $W_{\beta}[p]$ from the 
expansion of the exponentials
\begin{equation}
\label{eq:weightfactorbeta}	
W_{\beta}[p] \; = \; 
\prod_{x,\mu<\nu} \prod_{a,b,c,d}  \dfrac{ \left( \frac{\beta}{2} \right)^{p_{x,\mu\nu}^{abcd}}}{p_{x,\mu\nu}^{abcd}!}  \; ,
\end{equation}
and the weight factor $W_{H}[p]$ which collects the combinatorial factors from the Haar measure integral,
\begin{equation}
\label{eq:weightfactorhaar}	
W_{H}[p] \; = \; 
\prod_{x,\mu} \dfrac{\left(\frac{S_{x,\mu}^{11} + S_{x,\mu}^{22}}{2}\right)! 
\left(\frac{S_{x,\mu}^{12} + S_{x,\mu}^{21}}{2}\right)!}{\left(\frac{S_{x,\mu}^{11} + S_{x,\mu}^{22} + 
S_{x,\mu}^{12} + S_{x,\mu}^{21}}{2} + 1\right)!} \; .
\end{equation}
Here we make use of the evenness properties  (\ref{eq:Scombinations}) and write the weight factors already 
with factorials (compare (\ref{weightintegrals})).

In its dual form (\ref{eq:partitionsum4}) the partition function is a sum over configurations of cycle occupation numbers
$p_{x,\mu\nu}^{abcd} \in \mathbb{N}_{0}$ attached to the plaquettes $(x, \mu < \nu)$. At each link $(x,\mu)$
the $p_{x,\mu\nu}^{abcd}$ have to  obey constraints which are expressed in terms of the two Kronecker deltas
that relate components of the fluxes $J_{x,\mu}^{ab}$ at each link. We will discuss the geometrical interpretation of 
these constraints below. 

Each configuration comes with the weight factors $W_{\beta}[p]$ and $W_{H}[p]$ given in 
(\ref{eq:weightfactorbeta}) and (\ref{eq:weightfactorhaar}) which are both 
real and positive. Note, however, that the partition sum (\ref{eq:partitionsum4}) 
also contains the explicit sign factor $(-1)^{\sum_{x,\mu}J_{x,\mu}^{21}}$ which origins from the minus sign in 
the 2,1 matrix element in the parametrization (\ref{eq:parametrization}) of our SU(2) link variables. We will come
back to this sign after having discussed the geometrical meaning of the constraints.

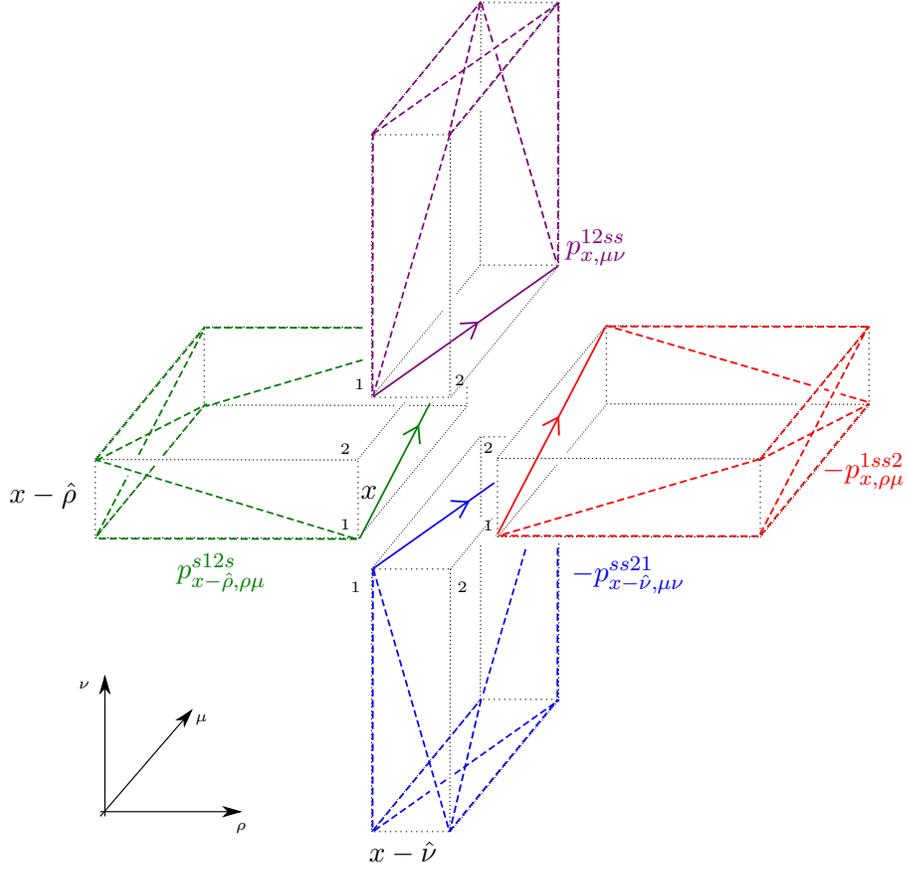
\begin{figure}[t!]
\vspace{5mm}
\begin{center}
\def\svgwidth{0.7\textwidth}
\hspace*{10mm}
%% Accompanies image file 'corrente12.pdf' (pdf, eps, ps)
\begingroup%
  \makeatletter%
  \providecommand\color[2][]{%
    \errmessage{(Inkscape) Color is used for the text in Inkscape, but the package 'color.sty' is not loaded}%
    \renewcommand\color[2][]{}%
  }%
  \providecommand\transparent[1]{%
    \errmessage{(Inkscape) Transparency is used (non-zero) for the text in Inkscape, but the package 'transparent.sty' is not loaded}%
    \renewcommand\transparent[1]{}%
  }%
  \providecommand\rotatebox[2]{#2}%
  \ifx\svgwidth\undefined%
    \setlength{\unitlength}{599.29966087bp}%
    \ifx\svgscale\undefined%
      \relax%
    \else%
      \setlength{\unitlength}{\unitlength * \real{\svgscale}}%
    \fi%
  \else%
    \setlength{\unitlength}{\svgwidth}%
  \fi%
  \global\let\svgwidth\undefined%
  \global\let\svgscale\undefined%
  \makeatother%
  \begin{picture}(1,0.98826961)%
    \put(0,0){\includegraphics[width=\unitlength]{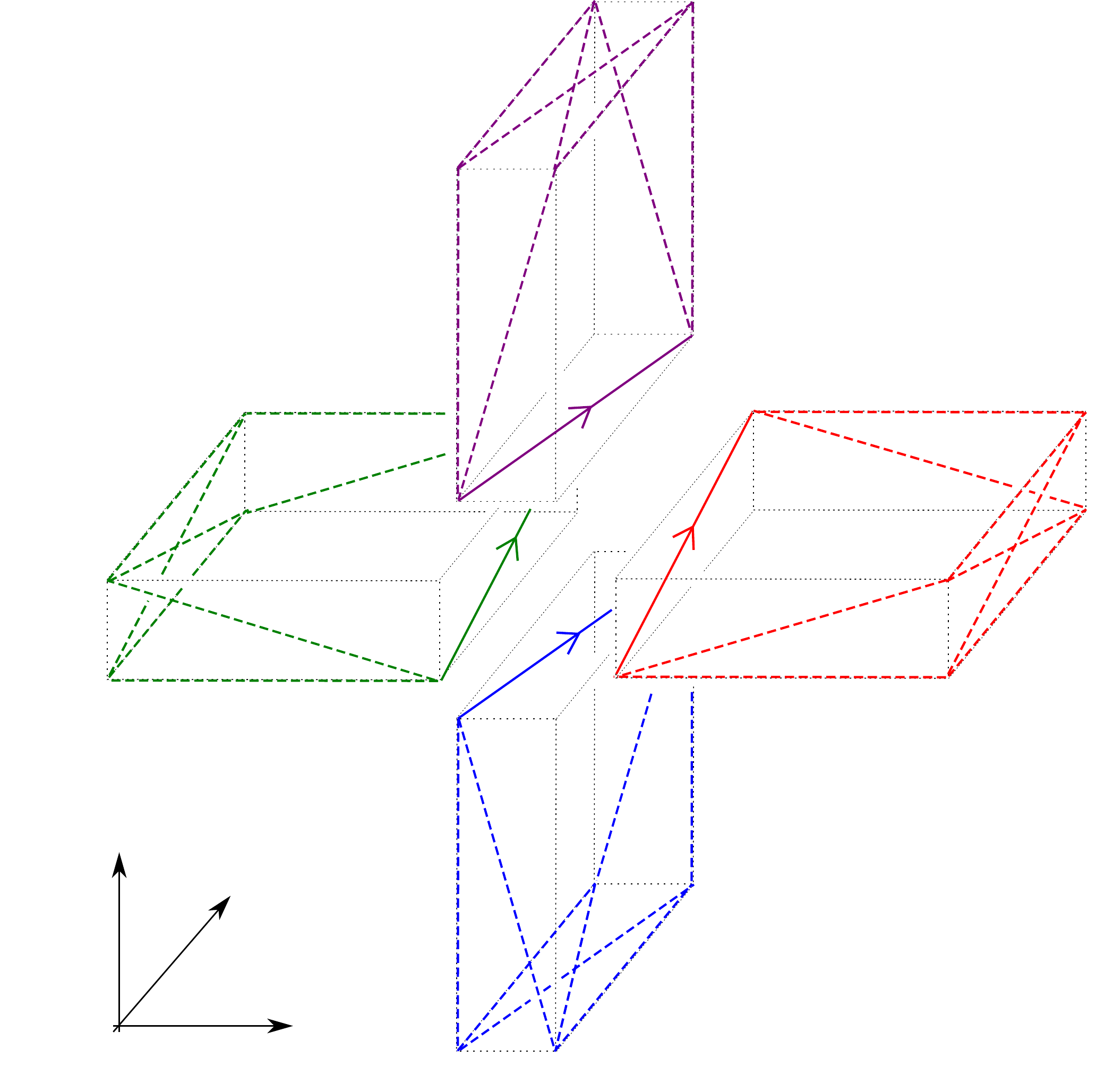}}%
    \put(0.2578568,0.03937669){\color[rgb]{0,0,0}\makebox(0,0)[lb]{\smash{\tiny$\rho$}}}%
    \put(0.07807751,0.20062075){\color[rgb]{0,0,0}\makebox(0,0)[lb]{\smash{\tiny$\nu$}}}%
    \put(0.21153366,0.16322944){\color[rgb]{0,0,0}\makebox(0,0)[lb]{\smash{\tiny$\mu$}}}%
    \put(0.38927314,0.31116376){\color[rgb]{0,0,0}\makebox(0,0)[lb]{\smash{\tiny$1$}}}%
    \put(0.51153331,0.31116376){\color[rgb]{0,0,0}\makebox(0,0)[lb]{\smash{\tiny$2$}}}%
    \put(0.41008122,0.00261242){\color[rgb]{0,0,0}\makebox(0,0)[lb]{\smash{$x - \hat{\nu}$}}}%
    \put(-0.00132968,0.41383362){\color[rgb]{0,0,0}\makebox(0,0)[lb]{\smash{$x - \hat{\rho}$}}}%
    \put(0.92939415,0.44215095){\color[rgb]{1,0,0}\makebox(0,0)[lb]{\smash{$-p^{1ss2}_{x,\rho\mu}$}}}%
    \put(0.64244529,0.3241621){\color[rgb]{0,0,1}\makebox(0,0)[lb]{\smash{$-p^{ss21}_{x-\hat{\nu},\mu\nu}$}}}%
    \put(0.19031199,0.32699383){\color[rgb]{0,0.50196078,0}\makebox(0,0)[lb]{\smash{$p^{s12s}_{x-\hat{\rho},\rho\mu}$}}}%
    \put(0.6358379,0.69889472){\color[rgb]{0.50196078,0,0.50196078}\makebox(0,0)[lb]{\smash{$p^{12ss}_{x,\mu\nu}$}}}%
    \put(0.40080404,0.41855318){\color[rgb]{0,0,0}\makebox(0,0)[lb]{\smash{$x$}}}%
    \put(0.37747427,0.38290099){\color[rgb]{0,0,0}\makebox(0,0)[lb]{\smash{\tiny$1$}}}%
    \put(0.54124282,0.38195707){\color[rgb]{0,0,0}\makebox(0,0)[lb]{\smash{\tiny$1$}}}%
    \put(0.39399273,0.54336585){\color[rgb]{0,0,0}\makebox(0,0)[lb]{\smash{\tiny$1$}}}%
    \put(0.54124282,0.47076809){\color[rgb]{0,0,0}\makebox(0,0)[lb]{\smash{\tiny$2$}}}%
    \put(0.37747427,0.46982414){\color[rgb]{0,0,0}\makebox(0,0)[lb]{\smash{\tiny$2$}}}%
    \put(0.50870158,0.54816876){\color[rgb]{0,0,0}\makebox(0,0)[lb]{\smash{\tiny$2$}}}%
  \end{picture}%
\endgroup%
\end{center}
\caption{Graphical illustration of the contributions from the cycle occupation numbers to the $J$-flux using the  
example of the $J_{x,\mu}^{12}$ element. For a description of the figure see the text.}
\label{fig:Jgeometry}
\end{figure}

The constraints (\ref{eq:constraints}) at each link $(x,\mu)$ connect components $J_{x,\mu}^{ab}$ of the $J$-fluxes 
defined in (\ref{eq:Jfluxes}). Thus for understanding the constraints we need to understand the geometrical 
interpretation of the fluxes $J_{x,\mu}^{ab}$. They are built from cycle occupation numbers $p_{x,\mu\nu}^{abcd}$
where two of the color indices are summed. Upon detailed inspection one finds that $J_{x,\mu}^{ab}$ is the sum 
over all cycle occupation numbers that live on those plaquettes which contain the link $(x,\mu)$ and lead from color 
$a$ to color $b$. The color indices for the other two corners are summed.  
The fluxes along the cycles are ordered, such that cycles where the link $(x,\mu)$ is run through in negative 
direction come with a negative sign in $J_{x,\mu}^{ab}$. 
In Fig.~\ref{fig:Jgeometry} we illustrate as an example the contributions to $J_{x,\mu}^{12}$: 
The central link $(x,\mu)$ is in the center of the plot, and we plot all plaquettes in the $\rho$-$\mu$
and $\mu$-$\nu$ planes that attach to $(x,\mu)$ with showing both their color layers. With full lines we 
show the flux that is kept fixed in $J_{x,\mu}^{12}$, 
i.e., the flux that connects the color index 1 at $x$ to the color index 2 at 
$x + \hat{\mu}$. This particular flux component
is contained in 4 different cycles in each of the attached plaquettes. These
cycles are summed over in the definition of $J_{x,\mu}^{ab}$ and with dashed lines we trace out the 
corresponding cycles.

\begin{figure}[t]
\vspace{5mm}
\begin{center}
\hspace*{5mm}
\includegraphics[scale=0.8,clip]{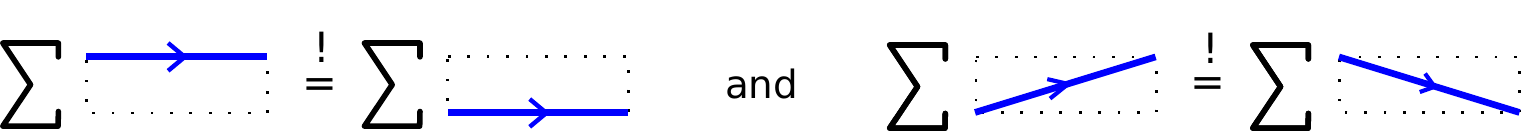}
\end{center}
\caption{Geometrical illustration of the two constraints Eq.(\ref{eq:constraints}) for the fluxes $J_{x,\mu}^{ab}$
on all links $(x,\mu)$.  The first constraint (lhs.~plot) requires the sum over all 1-1 flux to equal the sum over all 2-2
flux. The second constraint requires the sum over 1-2 fluxes to equal the sum over 2-1 fluxes.
\hfill}
\label{fig:fluxconservation}
\end{figure}

With the geometrical interpretation of the fluxes $J_{x,\mu}^{ab}$, we can now also interpret 
the two constraints given in (\ref{eq:constraints}): They imply that along all links $(x,\mu)$ the combined flux 
leading from color 1 to color 1 has to match the flux connecting 2 and 2, and that the flux from 1 to 2 has to 
match the flux from 2 to 1. For later use we introduce a geometrical illustration of the constraints, 
which we show in Fig.~\ref{fig:fluxconservation}.

The constraints also allow for a simple interpretation of the minus sign in the partition sum. A minus sign is
taken into account for every unit of  $J_{x,\mu}^{21}$ flux. Since by the constraint the $J_{x,\mu}^{21}$ flux equals the
$J_{x,\mu}^{12}$ flux which crosses the $J_{x,\mu}^{21}$ flux, the explicit 
minus sign in the partition sum simply takes the form
\begin{equation}
(-1)^{\sum_{x,\mu}J_{x,\mu}^{21}} \; = \; (-1)^{\, \# \, flux \; crossings} \; .
\end{equation} 
Having understood the geometrical interpretation of the constraints and the sign, 
simplifies the analysis of configurations considerably. 

It is easy to construct dual pure gauge 
configurations that obey all the constraints. Examples are closed orientable surfaces of plaquettes where certain cycle
occupation numbers are occupied. One can, e.g., start with all plaquette occupation numbers set to $0$,
and then sets $p_{x,\mu\nu}^{1111} = 1$ for all plaquettes of the surface. This choice guarantees that
all flux constraints on all links of the lattice are obeyed. 
Obviously such a configuration has no crossings of flux and thus a positive sign.
Subsequently one can choose a site in the surface and flip the values of the color indices on that site, 
i.e., one interchanges $1 \rightarrow 2$. This means that for all cycle occupation numbers that contain the site 
the corresponding color index is flipped. This changes the types of ACCs that are occupied on the surface, 
but it is easy to see that 
the constraints (\ref{eq:constraints}) remain intact. Repeating this for several (or all) 
sites in the surface generates more general closed surface configurations made of all 16 ACCs. 
Since flipping the color at a site and then changing the ACCs on the four plaquettes of the surface 
containing the site always changes the number of crossings by a multiple of 2, all these configurations have 
positive sign. This gives a large class of admissible configurations with positive sign. Generalizations of this
strategy to higher occupation numbers, surfaces that wind around the periodic boundaries, or to non-orientable 
surfaces are straightforward.

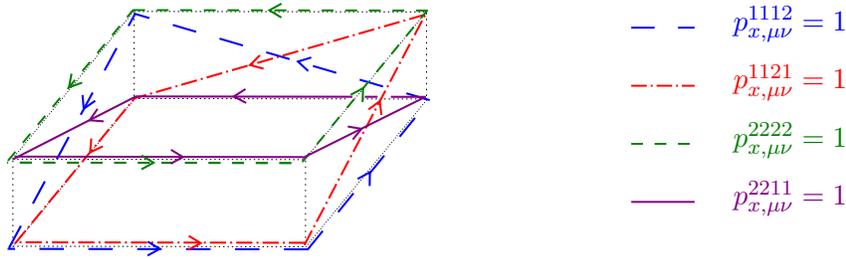
\begin{figure}[t!]
\centering
\def\svgwidth{0.7\textwidth}
\vskip5mm
%% Creator: Inkscape inkscape 0.48.4, www.inkscape.org
%% Accompanies image file 'unhappy.pdf' (pdf, eps, ps)
\begingroup%
  \makeatletter%
  \providecommand\color[2][]{%
    \errmessage{(Inkscape) Color is used for the text in Inkscape, but the package 'color.sty' is not loaded}%
    \renewcommand\color[2][]{}%
  }%
  \providecommand\transparent[1]{%
    \errmessage{(Inkscape) Transparency is used (non-zero) for the text in Inkscape, but the package 'transparent.sty' is not loaded}%
    \renewcommand\transparent[1]{}%
  }%
  \providecommand\rotatebox[2]{#2}%
  \ifx\svgwidth\undefined%
    \setlength{\unitlength}{538.25709386bp}%
    \ifx\svgscale\undefined%
      \relax%
    \else%
      \setlength{\unitlength}{\unitlength * \real{\svgscale}}%
    \fi%
  \else%
    \setlength{\unitlength}{\svgwidth}%
  \fi%
  \global\let\svgwidth\undefined%
  \global\let\svgscale\undefined%
  \makeatother%
  \begin{picture}(1,0.29057908)%
    \put(0,0){\includegraphics[width=\unitlength]{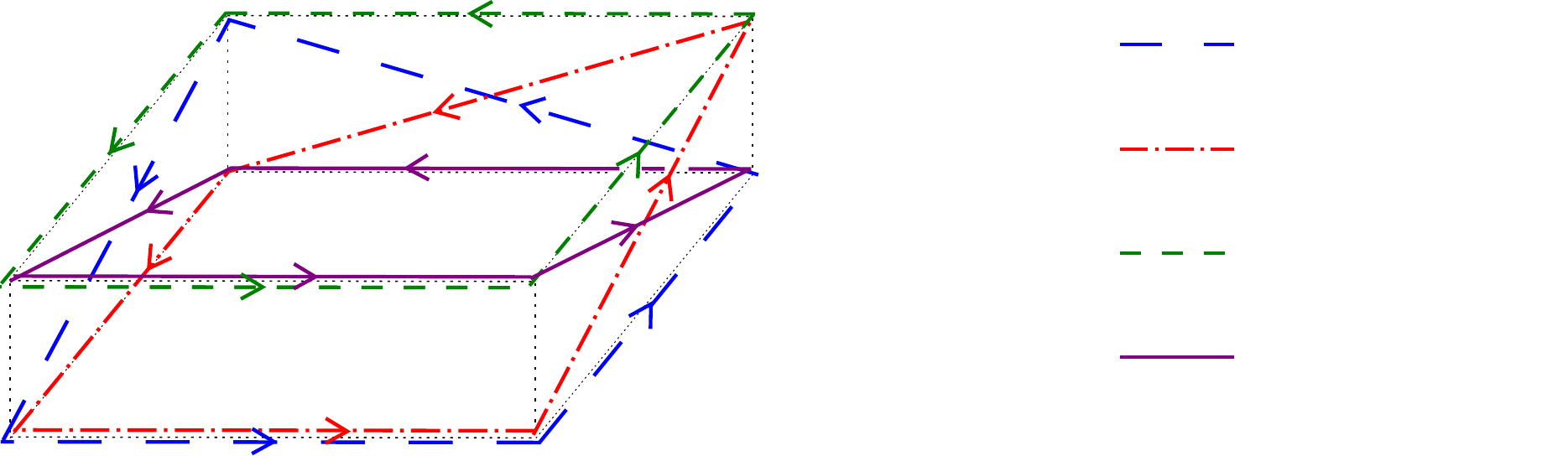}}%
    \put(0.83120867,0.25875274){\color[rgb]{0,0,1}\makebox(0,0)[lb]{\smash{$p^{1112}_{x,\mu\nu}=1$}}}%
    \put(0.83120867,0.19221858){\color[rgb]{1,0,0}\makebox(0,0)[lb]{\smash{$p^{1121}_{x,\mu\nu}=1$}}}%
    \put(0.83120867,0.12568436){\color[rgb]{0,0.50196078,0}\makebox(0,0)[lb]{\smash{$p^{2222}_{x,\mu\nu}=1$}}}%
    \put(0.83120867,0.05915015){\color[rgb]{0.50196078,0,0.50196078}\makebox(0,0)[lb]{\smash{$p^{2211}_{x,\mu\nu}=1$}}}%
  \end{picture}%
\endgroup%
\caption{Example of a configuration that contributes to the partition function with a negative sign. We show
the four cycle occupation numbers that are set to 1. Obviously all constraints are obeyed. The configuration has three
flux crossings and thus a negative sign. \hfill}
\label{fig:unhappy}
\end{figure}

However, configurations with negative sign are not excluded completely by the constraints. The lowest negative sign 
configuration which one can construct appears at order $\beta^4$, i.e., it has four non-zero 
cycle occupation numbers. An example is, e.g., given by 
\begin{equation}
p_{x,\mu\nu}^{1112} = 1 \; , \; p_{x,\mu\nu}^{1121} = 1 \; , \;
p_{x,\mu\nu}^{2222} = 1 \; , \; p_{x,\mu\nu}^{2211} = 1 \; .
\end{equation}
This configuration is shown in Fig.~\ref{fig:unhappy}. It obviously obeys all the constraints, but 
has an odd number of crossings such that it contributes to the partition function with a negative sign.

Before we come to generalizing the dual representation with ACCs to the case of SU(2) gauge fields coupled to fermions, 
we briefly discuss observables in pure gauge theory and their representation in the dual formulation of the theory. 
The simplest observable obviously
is the plaquette expectation value $\langle U_p \rangle \equiv \frac{1}{6V} \, \partial  \ln Z / \partial (\beta/2) =
\frac{1}{6V} \frac{1}{Z} \partial Z / \partial (\beta/2)$,
where $V$ denotes the total number of lattice sites. It is straightforward to compute the derivative of $Z$ with 
respect to $\beta/2$ also for the dual partition sum (\ref{eq:partitionsum4}). One easily finds the result
\begin{equation}
\langle U_p \rangle \, = \, \frac{1}{6V} \frac{1}{Z} \frac{2}{\beta}
\left\langle \sum_{x,\mu<\nu} \sum_{a,b,c,d}  p_{x,\mu\nu}^{abcd} \right \rangle \; ,
\end{equation}
where the vacuum expectation value on the right hand side is now understood in terms of the dual variables. Thus the 
representation of the plaquette expectation value in the dual formulation is the sum over all cycle occupation numbers. In
a similar way one can compute higher derivatives with respect to $\beta/2$ to obtain higher moments of the plaquette. In the 
dual language they correspond to higher moments of the sum over cycle occupation numbers. 

A simple generalization leads to the dual representation of correlators of plaquettes: We can introduce a different 
gauge coupling $\beta_{x,\mu\nu}$ for each plaquette $(x, \mu < \nu)$. The steps of the dualization go through in 
exactly the same way and we find that the couplings $\beta_{x,\mu\nu}$ enter the partition sum in the form 
\begin{equation}
\prod_{x,\mu<\nu} \left( \frac{\beta_{x,\mu\nu}}{2} \right)^{\sum_{a,b,c,d}  p_{x,\mu\nu}^{abcd}} \; .
\label{betalocal}
\end{equation}
Correlators of plaquettes can now be obtained as derivatives, $\partial^2 \ln Z / \partial (\beta_{x,\mu \nu}/2) \partial
(\beta_{x^\prime,\mu^\prime \nu^\prime}/2)$, evaluated at $\beta_{x,\mu \nu} = \beta$. From (\ref{betalocal}) 
one finds their dual representation in terms of correlators of $\sum_{a,b,c,d}  p_{x,\mu\nu}^{abcd}$.

Finally we comment on how to represent the Wilson loop in the dual representation. The Wilson loop is the trace
of the product of gauge links around a closed contour $C$. We can write the matrix product and the trace with explicit
sums over color indices and obtain,
\begin{equation}
\mbox{Tr} \, \prod_{l \in C} U_l \; = \; \sum_{a_1,a_2, \, ... \, a_n} 
U_{l_1}^{\,a_1 a_2} \, U_{l_2}^{\,a_2 a_3} \, ..... \, U_{l_n}^{\,a_n a_1} \; .
\label{wilsonloop}
\end{equation}
Here the $U_l$ denote the gauge variables on the links $l$ of the contour $C$ which we assume to consist of
$n$ links. For links $l$ that are run through in negative direction the hermitian conjugate link variable is used. Thus
we can write the Wilson loop as a sum of closed paths in color space along the contour $C$. Each of these paths 
introduces color flux along the links of the paths on the two-layer lattice. This flux 
has to be compensated by activating cycle occupation numbers 
on a surface that has the contour $C$ as its boundary, such that all constraints are obeyed, as illustrated in 
Fig.~\ref{fig:fluxconservation}. This is exactly the same structure as one finds in the case of fermion loops
which we discuss next.

\section{Adding fermionic matter}

After having introduced the concept of ACCs for pure gauge theories we now show that the corresponding 
techniques can also be implemented for gauge theories with fermions. We here consider staggered 
fermions which in a path integral are represented 
by Grassmann variables $\psi^a_x$ and $\overline{\psi}^a_x$. Here $x$ denotes the space-time index and $a$ the 
color index. The fermions have periodic boundary conditions for the spatial direction and anti-periodic boundary
conditions for the temporal direction, i.e., the $\mu = 4$ direction. 

The fermionic partition sum $Z_F[U]$ in a background gauge field configuration is given by 
\begin{equation}
Z_F[U] \; = \; \int \!\! D[\, \overline{\psi}, \psi] \; e^{-S_F[\,\overline{\psi}, \psi,U]} \; ,
\end{equation}
where the measure is a product over Grassmann measures, 
$D[\, \overline{\psi}, \psi] = \prod_{x,a} d \psi^a_x d \overline{\psi}^a_x$. The full partition sum $Z$ is then obtained as 
$Z = \int D[U]  e^{-S_G[U]} Z_F[U]$.
The staggered action in a SU(2) background field is given by 
\begin{eqnarray}
S_F[\,\overline{\psi}, \psi,U]  & = &  \sum_x \Big[ m \, \overline{\psi}_x \psi_x + \sum_{\mu} \frac{\gamma_{x,\mu}}{2}
\Big( \, \overline{\psi}_x U_{x,\mu} \psi_{x + \hat\mu} - \overline{\psi}_{x+\hat\mu} U_{x,\mu}^\dagger \psi_{x} 
\Big) \Big] 
\nonumber \\
& = &\sum_x \Big[ \, m \!\sum_{a} 
\overline{\psi}_x^{\,a} \psi_x^{a} + \sum_{\mu} \frac{\gamma_{x,\mu}}{2} \sum_{a,b}
\Big( \, \overline{\psi}^{\,a}_x U_{x,\mu}^{\,ab} 
\psi_{x + \hat\mu}^{b} - \overline{\psi}_{x+\hat\mu}^{\,b} U_{x,\mu}^{\, ab \, \star} \psi_{x}^{a} 
\Big) \Big] \; ,
\label{fermionaction}
\end{eqnarray}
where in the first line we use matrix/vector notation for gauge links and fermions, while in the second line
the sums over the color indices are made explicit. The staggered sign factors are given by
\begin{equation}
\gamma_{x,1} = 1\; , \; \gamma_{x,2} = (-1)^{x_1} \; , \; 
\gamma_{x,3} = (-1)^{x_1+x_2} \; , \; \gamma_{x,4} = (-1)^{x_1+x_2+x_3} \; .
\end{equation}
Since all terms in the fermion action are bilinears in 
Grassmann variables they commute with each other and we can write the exponential of the action as a 
product over individual exponentials. We find for the fermionic partition sum,
\begin{eqnarray}
\hspace*{-6mm} && Z_F[U] \, = \int \!\!\! D[\, \overline{\psi}, \psi]  \prod_x \prod_a e^{-m \overline{\psi}_x^{\,a} \psi_x^a} \; 
\prod_{x,\mu} \prod_{a,b} e^{- \,\frac{\gamma_{x,\mu}}{2} \overline{\psi}^{\,a}_x U_{x,\mu}^{\,ab} \psi_{x + \hat\mu}^b} \;
e^{ \, \frac{\gamma_{x,\mu}}{2} \overline{\psi}_{x+\hat\mu}^{\,b} U_{x,\mu}^{\, ab\, \star} \psi_{x}^a} 
\nonumber \\ 
\hspace*{-6mm} && = \int \!\!\!D[\, \overline{\psi}, \psi]  \prod_x \prod_a \sum_{s_x^a = \, 0}^{1} \! 
( -m \overline{\psi}_x^{\,a} \psi_x^a )^{s_x^a} 
% \nonumber \\
% &&  \times
\prod_{x,\mu} \prod_{a,b} 
\sum_{k_{x,\mu}^{\;ab} = \, 0 }^{1} \!\!\! 
(- \frac{\gamma_{x,\mu}}{2} \, \overline{\psi}^{\,a}_x U_{x,\mu}^{\,ab} \psi_{x + \hat\mu}^b)^{k_{x,\mu}^{\; ab}} \!
\sum_{ \overline{k}_{x,\mu}^{\;ab} = \, 0}^{1}\!\!\! 
( \frac{\gamma_{x,\mu}}{2} \, \overline{\psi}_{x+\hat\mu}^{\,b} U_{x,\mu}^{\, ab\, \star} 
\psi_{x}^a)^{\overline{k}_{x,\mu}^{\; ab}}
\nonumber \\
\hspace*{-6mm} &&= \, \frac{1}{2^{2V}} \! \sum_{\{s,k,\overline{k}\}} \!\! (2m)^{\, \sum_{x,a} s_x^a} \; \prod_{x,\mu} \prod_{a,b} 
( U_{x,\mu}^{\,ab} )^{k_{x,\mu}^{\; ab}} \, 
( U_{x,\mu}^{\, ab\, \star} )^{\overline{k}_{x,\mu}^{\; ab}}  \; (-1)^{k_{x,\mu}^{\; ab}} \; 
( \, \gamma_{x,\mu}\,) ^{k_{x,\mu}^{\; ab} + \overline{k}_{x,\mu}^{\; ab} } 
\nonumber \\
\hspace*{-6mm} && \hspace{40mm} \times \!\!
\int \!\! D[\, \overline{\psi}, \psi] \! \prod_x \prod_a  (\overline{\psi}_x^{\,a} \psi_x^a )^{s_x^a} \prod_{x,\mu} \prod_{a,b}
(\overline{\psi}^{\,a}_x \psi_{x + \hat\mu}^b)^{k_{x,\mu}^{\; ab}} 
(\overline{\psi}_{x+\hat\mu}^{\,b} \psi_{x}^a)^{\overline{k}_{x,\mu}^{\; ab}} \; .
\label{fermiondual1}
\end{eqnarray}
In the second line we have Taylor-expanded the individual potentials of the Grassmann bilinears. Due to nilpotency 
of the Grassmann variables these Taylor series terminate after the linear term. Note that we introduce an expansion
index for each bilinear in the action: $s_x^a  = 0,1$ is used for the color components
$a$ in the mass terms (we will refer to these also as {\sl ''monomers''}), 
$k_{x,\mu}^{\,ab} = 0,1$ for the forward hopping terms that connect colors $a$ and $b$ and  
$\overline{k}_{x,\mu}^{\,ab} = 0,1$ for the corresponding backward hopping. These expansion indices
will be the dual variables for the fermions.  We remark at this point that in the last step 
we dropped the minus sign in front of the mass $m$. We will see below that the number of monomers is even,
i.e., the exponent $\sum_{x,a} s_x^a$ is a multiple of 2 and thus this sign is irrelevant and can be omitted. 

\begin{figure}[t]
\begin{center}
\vskip5mm
\includegraphics[scale=0.53,clip]{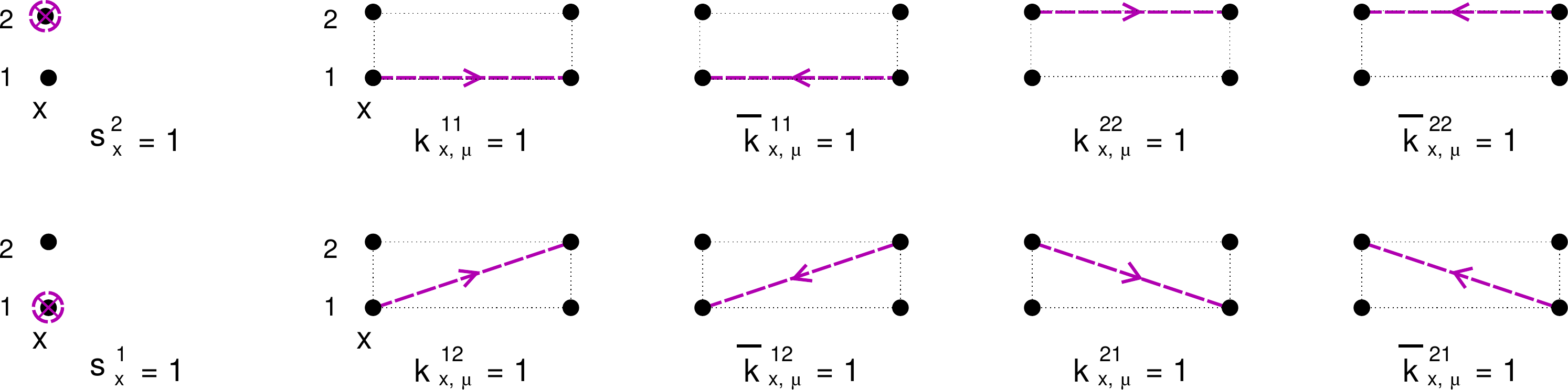}
\end{center}
\caption{Graphical representation of the dual variables for the fermions. 
The first two diagrams on the very left represent the monomers $s_x^a$, 
the arrows are for the dual variables $k_{x,\mu}^{\,ab}, \overline{k}_{x,\mu}^{\,ab}$.}
\label{fig:grassmann_dual}
\end{figure}

The dual variables for the fermions can be represented graphically in a way equivalent to the ACCs for the gauge fields. 
In Fig.~\ref{fig:grassmann_dual} we show the corresponding graphical elements. The monomers $s_x^a$ 
activate a term $\overline{\psi}_x^{\,a} \psi_x^a$ which completely saturates the Grassmann integral at a site (see below)
and is independent of the gauge fields. We represent them by a circle around the corresponding site and color 
(the two diagrams on the very left of Fig.~\ref{fig:grassmann_dual}). The dual variables for forward 
hopping $k_{x,\mu}^{\,ab}$ activate a nearest neighbor term of the Grassmann variables which comes with the
matrix element $U_{x,\mu}^{\,ab}$ and thus is represented by an arrow that points in positive $\mu$-direction. 
According to the different choices for $a$ and $b$ we have 4 different possibilities to connect the color indices. 
The dual variable $\overline{k}_{x,\mu}^{\,ab}$ activates backward hopping and comes with the
matrix element $U_{x,\mu}^{\, ab\, \star}$. Consequently we represent it with an arrow in negative direction,
which again can combine the colors on the neighboring sites in four different ways.
 
In the last step of Eq.~(\ref{fermiondual1}) we reorganize the terms: We introduce the notation 
$\sum_{\{s,k,\overline{k}\}}$ for the sum over all possible values of the expansion indices and collect all
factors (including overall factors of $1/2$) that do not depend on the Grassmann 
variables in front of the Grassmann integral. Among these factors are also pure sign factors: the terms 
$(-1)^{k_{x,\mu}^{\; ab}}$ introduce a minus sign for every activated forward hopping term ($k_{x,\mu}^{\; ab} = 1$).
The factors $( \, \gamma_{x,\mu}\,) ^{k_{x,\mu}^{\; ab} + \overline{k}_{x,\mu}^{\; ab} }$ collect the signs from 
the staggered sign factors.  

The Grassmann integral in the very last line gives either $0$ or $\pm 1$, depending on the values of the 
dual variables $s_x^a, k_{x,\mu}^{\,ab}$ and $\overline{k}_{x,\mu}^{\,ab}$. A Grassmann integral is non-vanishing
only when each Grassmann variable $\psi^a_x, \overline{\psi}^a_x$ appears exactly once\footnote{In the following
we use the terminology ''the Grassmann integral is saturated'' for this condition.}. The dual
variables $s_x^a, k_{x,\mu}^{\,ab}$ and $\overline{k}_{x,\mu}^{\,ab}$ activate (dual variable = 1) or deactivate 
(dual variable = 0) the corresponding Grassmann variables and thus for a non-zero result of the Grassmann integral 
the dual variables have to obey constraints. 

We stress at this point, that the color index $a = 1,2$ of the Grassmann variables  $\psi^a_x$ and 
$\overline{\psi}^a_x$ can be interpreted on the same footing as the space-time index $x$ that labels the lattice sites. 
Again we can interpret the two values $a=1,2$ as labels for two layers of the 4-dimensional
lattice labelled by $x$. Monomer terms $\overline{\psi}_x^{\,a} \psi_x^a$ live on only one site of one of the two layers. 
Forward hopping terms $\overline{\psi}^{\,a}_x \psi_{x + \hat\mu}^b$ and backward hopping terms
$\overline{\psi}_{x+\hat\mu}^{\,b} \psi_{x}^a$ can 
either connect neighboring sites on the same layer ($a = b$), or connect neighboring sites on different layers
($a \neq b$). A term that connects layer 1 and layer 2 at the same site $x$ does not exist. 

The configurations that saturate the Grassmann integrals on all sites of the two layers have a simple geometrical 
interpretation in terms of monomers, dimers and loops. The simplest choice for saturating the Grassmann 
integral on a site $x$ in layer $a$ is to activate a monomer $\overline{\psi}^a_x \psi^a_x$ by setting the 
corresponding $s_x^a = 1$. Since the monomer dual variables $s_x^a$ do not give rise to explicit signs and 
the Grassmann variables are already in the canonical order for the Grassmann integral ($\overline{\psi}^a_x$ to 
the left of the corresponding $\psi^a_x$), a monomer contributes a factor of  $+1$. 

\begin{figure}[t]
\begin{center}
\vskip5mm
\includegraphics[scale=0.53,clip]{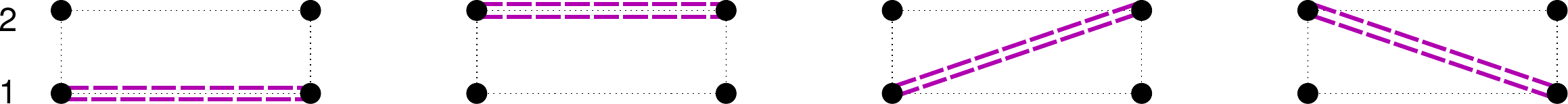}
\end{center}
\caption{Graphical representation of the possible dimers. Dimers saturate the Grassmann integrals for one of
the two colors of two neighboring sites. Four different color combinations are possible.}
\label{fig:dimers}
\end{figure}

Another simple choice is to place dimers by setting $k_{x,\mu}^{\,ab} =  \overline{k}_{x,\mu}^{\,ab} = 1$. This
dimer connects the color $a$ at $x$ with the color $b$ at $x+\hat{\mu}$. In the Grassmann integral this choice
activates the terms 
\begin{equation}
\overline{\psi}^{\,a}_x \, \psi_{x + \hat\mu}^b \, \overline{\psi}_{x+\hat\mu}^{\,b}  \, \psi_{x}^a \; = \; 
- \; \overline{\psi}^{\,a}_x  \, \psi_{x}^a \, \overline{\psi}_{x+\hat\mu}^{\,b} \, \psi_{x + \hat\mu}^b \; ,
\label{dimer}
\end{equation}
which in the second step we have brought into the canonical order such that the corresponding Grassmann 
integral gives $+1$. From the necessary interchanges of the Grassmann variables we have picked up a 
minus sign. However, this minus sign is compensated by the explicit minus sign for the forward hop 
(i.e., $(-1)^{k_{x,\mu}^{\,ab}} = (-1)^1 = -1$). The staggered sign factors do not contribute an additional sign
($( \, \gamma_{x,\mu}\,)^{k_{x,\mu}^{\; ab} + \overline{k}_{x,\mu}^{\; ab} } = ( \, \gamma_{x,\mu}\,)^2 = 1$), 
and we conclude that the dimers saturate the Grassmann integrals at the endpoints $x,a$ and $x+\hat{\mu}, b$,
and, like the monomers, contribute a factor of $+1$. In Fig.~\ref{fig:dimers} we illustrate the four different dimers that 
are possible.

The third possibility to saturate the Grassmann integral is to build an oriented non-selfintersecting 
loop $L$ out of hopping terms, such that for  a link $(x, \mu)$ 
that is run through in positive direction we set $k_{x,\mu}^{\,ab} = 1$ and for a link run through in negative
direction we set $\overline{k}_{x,\mu}^{\; ab} = 1$. Note that the loops can either stay in the same color layer ($a = b$),
or hop between the layers ($a \neq b$). 
The loops come with non-trivial signs which we will discuss below.

We conclude that the Grassmann integral leads to a constraint for the 
configurations of the monomer- ($s_x^a$) and link- ($k_{x,\mu}^{\; ab}, \overline{k}_{x,\mu}^{\; ab}$) activation indices,
restricting admissible configurations to those configurations where each site of the lattice with the two color layers is 
either occupied by a monomer, is the endpoint of a dimer or is run through by a loop. Dimers saturate two sites of the lattice 
and loops an even number of sites (assuming that all lattice extensions are even). Thus the number of monomers is 
even as we already remarked above. 

While the monomers and dimers come only with positive signs, for each loop $L$ we need to calculate the
overall sign which comes from several sources:
We have to take into account an explicit minus sign for each loop from bringing 
the Grassmann variables into canonical order, equivalent to the sign which we picked up in (\ref{dimer}) when reordering
the Grassmann variables in the dimer. Furthermore, since we use anti-periodic temporal boundary conditions, loops that 
wind an odd number of times across the compact temporal direction pick up a minus sign.  
This sign is given by $(-1)^{W_L}$, where $W_L$ is the total winding number of the loop $L$ around compact time. 

The loops pick up further signs, one of them from the factors $\prod_{x,\mu,a,b} (-1)^{k_{x,\mu}^{\; ab}}$. 
This counts the minus signs which we 
have to take into account for each forward hop. For trivially closing loops the number of forward hops is exactly 
half of the hops in a loop, i.e., half of the links the loop is made of. If we define the length $|L|$ of a loop $L$ by the
number of hops it consists of, then for trivially closing loops the sign from $\prod_{x,\mu,a,b} (-1)^{k_{x,\mu}^{\; ab}}$ 
can be written as
$(-1)^{|L|/2}$. Loops that wind around the compact time or space direction do not share the property of equal numbers of 
forward and backward hops and one would have to distinguish different cases. For simplicity we here assume that  the 
temporal and spatial extents of the lattice are all multiples of 4 and it is easy to see, that then $(-1)^{|L|/2}$
always correctly takes into account the signs from forward hopping.

Finally, the loops also obtain the signs 
$\prod_{x,\mu,a,b} ( \, \gamma_{x,\mu}\,) ^{k_{x,\mu}^{\; ab} + \overline{k}_{x,\mu}^{\; ab} }$ 
from the staggered sign factors. For each hop of the loop 
along a link $(x,\mu)$ a factor $\gamma_{x,\mu}$ is taken into account, independent of whether the loop runs through 
$(x,\mu)$ in positive direction ($k_{x,\mu}^{\; ab} = 1$) or in negative direction ($\overline{k}_{x,\mu}^{\; ab} = 1$). 
We also stress that the way a loop moves in color space, i.e., the values of $a$ and $b$ on the activated 
$k_{x,\mu}^{\; ab}, \overline{k}_{x,\mu}^{\; ab}$, is irrelevant for the sign coming from the staggered factors. 

Actually, for many loops the sign from the staggered factors can be expressed in a rather simple way: 
We first note the trivial
property,
\begin{equation}
\sigma_{x,\mu\nu} \; = \; 
\gamma_{x,\mu} \,  \gamma_{x+\hat{\mu},\nu} \, \gamma_{x+\hat{\nu},\mu} \, \gamma_{x,\nu} \; = \; -\, 1 \;,
\end{equation}
i.e., the product $\sigma_{x,\mu\nu}$ 
of the four staggered factors for the links of a plaquette $(x,\mu<\nu)$ is always equal to $-1$. If we 
now multiply two of these products around neighboring plaquettes, then the staggered factor
along the common link gets squared and drops out. What remains is the product of the staggered factors along the 
boundary of the surface constituted by the two plaquettes. This procedure can be iterated and for arbitrary 
surfaces that are made of plaquettes we obtain the product of the staggered sign factors along the loop 
that is the boundary of that surface as the product of the corresponding plaquette factors $\sigma_{x,\mu\nu}$. 
For a surface made of $P$ plaquettes we thus have a factor $(-1)^P$ which is the result for the product of
staggered factors of the loop that is the boundary of that surface. Note that we here deal with loops embedded in four 
dimensions, where all loops are isotopic to $S^1$ (see, e.g., \cite{Ranicki}), i.e., they are free of knots and the surfaces
they span are isotopic to discs, such that our procedure of building up the staggered sign from the plaquette factors 
$\sigma_{x,\mu\nu}$ is always applicable. 

The surface that has a given loop $L$ as its
boundary is of course not unique. However, the numbers of plaquettes that are needed for spanning different surfaces 
with the same loop $L$ as their boundary differ by multiples of 2, such that for all loops that are the boundaries 
of surfaces made of plaquettes we find the simple expression $(-1)^{P_L}$ for the product of staggered factors along the
loops. Here $P_L$ is the number of plaquettes in an arbitrary surface that has $L$ as its boundary. 
Note that also non-orientable surfaces with a boundary such as a M\"obius strip can be constructed with 
plaquettes such that the result $(-1)^{P_L}$ holds also for their boundary loops. Also self intersection of 
surfaces does not change this result.

Not all closed loops on our 4-torus are the boundaries of surfaces, in particular loops that wind around
one of the compact directions. However, also for these loops one can compute the product of staggered factors
in a simple way. Let us start with the example of a straight loop that winds around the compact time direction ($\mu = 4$)
and is located at spatial position $\vec{x}$. The corresponding product of staggered factors is
\begin{equation}
\prod_{x_4 = 0}^{N_4-1} \gamma_{\;(\vec{x},x_4),4} \; = \; (\, \gamma_{\;(\vec{x},0),4} \,)^{N_4} \; = \; + 1 \; ,
\end{equation}
where in the first step we have used the fact that $\gamma_{(\vec{x},x_4),4}$ is independent of $x_4$ and in the
second step assumed that the number $N_4$ of lattice sites in 4-direction is even (actually we have already assumed
above that it
is a multiple of $4$). In exactly the same way one finds that the product of staggered sign factors is equal to
1 for all straight loops that wind around any of the four compact directions, spatial or temporal. 
These loops can now be deformed by 
attaching plaquettes and the corresponding factors $\sigma_{x,\mu\nu}$ to obtain a more general winding loop. As before,
the staggered factors on the links where we attach plaquettes get squared and cancel, such that we obtain the results
for the product of staggered factors of a general winding loop $L$ as $(-1)^{P_L}$, where now $P_L$ is the number of 
plaquettes needed to generate the loop from a straight winding loop. 

Thus we can summarize our discussion of the sign factor $\mbox{sign} \,(L)$ for a loop $L$ by the following formula:
\begin{equation}
\mbox{sign}\,(L) \; = \; (-1)^{ \, |L|/2 \, + \, W_L \, + \,  P_L \, + \, 1 } \; ,
\label{loopsign}
\end{equation}
where $|L|$ is the length of the loop, $W_L$ the number of windings around the compact time direction, $P_L$ the 
number of plaquettes of a surface with $L$ as its boundary (or for spatially or temporally winding loops the numbers of 
plaquettes in the surface between the loop and a straight loop with the same winding).

Putting things together we find the following expression for the fermionic partition sum,
\begin{equation}
Z_F[U] \, =   \, 
\frac{1}{2^{2V}} \! \sum_{\{s,k,\overline{k}\}} \!\! \!\! C_{F}[s,k,\overline{k}\,] \; W_M[s] \; \prod_{L} \, \mbox{sign} \,(L) 
 \prod_{x,\mu} \prod_{a,b} 
( U_{x,\mu}^{\,ab} )^{k_{x,\mu}^{\; ab}} \, 
( U_{x,\mu}^{\, ab\, \star} )^{\overline{k}_{x,\mu}^{\; ab}}  \; ,
\label{fermiondual2}
\end{equation}
where we introduced the monomer weights 
$W_M[s] = (2m)^{\, \sum_{x,a} s_x^a}$. The fermion constraint $C_{F}[s,k,\overline{k}\,]$ is 1
only for admissible configurations of the dual variables $s_x^{a}, k_{x,\mu}^{\; ab}$ and
$\overline{k}_{x,\mu}^{\; ab}$ where each site of the double layer lattice is either occupied by a monomer, 
is the endpoint of a dimer, or is run through by a loop. Otherwise we have $C_{F}[s,k,\overline{k}\,] = 0$.
Each loop comes with a sign given by (\ref{loopsign}) and the last product collects all link matrix elements
that are activated along the loops. These link matrix elements still have to be integrated over when computing
the full partition sum. 

The full partition sum is obtained by integrating $e^{-S_G[U]} \, Z_F[U] $ with the product Haar measure $D[U]$,
\begin{eqnarray}
Z & = & \int \!  D[U]  e^{-S_G[U]} Z_F[U] 
\nonumber \\
& = & \frac{1}{2^{2V}} \!\! \sum_{\{s,k,\overline{k}\}} \!\! \!\! C_{F}[s,k,\overline{k}\,] \; W_M[s] \; 
\prod_{L} \, \mbox{sign} \,(L) 
\int \!  D[U]  e^{-S_G[U]}  \prod_{x,\mu} \prod_{a,b} 
( U_{x,\mu}^{\,ab} )^{k_{x,\mu}^{\; ab}} \, 
( U_{x,\mu}^{\, ab\, \star} )^{\overline{k}_{x,\mu}^{\; ab}}  \; .
\label{fermiondual3}
\end{eqnarray}
With the techniques of the previous section it is now straightforward to compute the integral over the gauge 
links which we have collected in the end of (\ref{fermiondual3}). It is exactly the same integral as we solved for the 
pure gauge theory case with the additional insertion of the factors 
$\prod_{x,\mu} \prod_{a,b} ( U_{x,\mu}^{\,ab} )^{k_{x,\mu}^{\; ab}} \, 
( U_{x,\mu}^{\, ab\, \star} )^{\overline{k}_{x,\mu}^{\; ab}}$.  Exactly the same integrands appear in the intermediate result 
(\ref{eq:partitionsum2}), but with the powers $N_{x,\mu}^{\; ab}$ and $\overline{N}_{x,\mu}^{\; ab}$. To obtain the 
dualization with the additional insertions from the fermion loops we simply need to replace 
$N_{x,\mu}^{\; ab}$ by $N_{x,\mu}^{\; ab}+k_{x,\mu}^{\; ab}$ and $\overline{N}_{x,\mu}^{\; ab}$ by 
$\overline{N}_{x,\mu}^{\; ab}+\overline{k}_{x,\mu}^{\; ab}$ and repeat the steps of Section 2. 
This implies that also the weights $W_H [p]$ which collect the combinatorial factors coming from
the Haar measure integration change. In addition to the cycle occupation numbers
$p_{x,\mu\nu}^{abcd}$ they now also depend on the dual variables $k_{x,\mu}^{\; ab}$ and 
$\overline{k}_{x,\mu}^{\; ab}$ which activate forward and backward fermionic hops,
\begin{equation}
\label{weighthaarfinal}
W_H [p,k,\overline{k}]  \; = \; 
\prod_{x,\mu} \dfrac{\!\left(\!\frac{S_{x,\mu}^{11} +k_{x,\mu}^{\; 11}+\overline{k}_{x,\mu}^{\; 11}+ 
S_{x,\mu}^{22} +k_{x,\mu}^{\; 22}+\overline{k}_{x,\mu}^{\; 22}}{2}\!\right) ! \; 
 \left(\frac{S_{x,\mu}^{12} +k_{x,\mu}^{\; 12}+\overline{k}_{x,\mu}^{\; 12} + 
S_{x,\mu}^{21}+k_{x,\mu}^{\; 21}+\overline{k}_{x,\mu}^{\; 21}}{2}\right) !}{\!\left(\frac{
S_{x,\mu}^{11} + k_{x,\mu}^{\; 11}+\overline{k}_{x,\mu}^{\; 11} + 
S_{x,\mu}^{22} + k_{x,\mu}^{\; 22}+\overline{k}_{x,\mu}^{\; 22} + 
S_{x,\mu}^{12} + k_{x,\mu}^{\; 12}+\overline{k}_{x,\mu}^{\; 12} + 
S_{x,\mu}^{21} + k_{x,\mu}^{\; 21}+\overline{k}_{x,\mu}^{\; 21}  }{2} + 1\!\right) !} \; .
\end{equation}

\begin{figure}[t]
\begin{center}
\vskip5mm
\includegraphics[scale=0.8,clip]{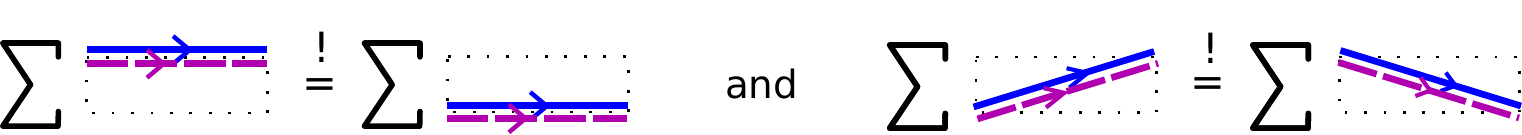}
\end{center}
\caption{Geometrical illustration of the generalized constraints that combine
the gauge fluxes $J_{x,\mu}^{ab}$ (full lines) and the matter fluxes $k_{x,\mu}^{\,ab} - \overline{k}_{x,\mu}^{\,ab}$
(dashed lines). Similar to the pure gauge case shown in Fig.~\ref{fig:fluxconservation} the sum over all 1-1 
fluxes must equal the sum over all 2-2 fluxes and the sum over 1-2 fluxes must equal the sum over 2-1 fluxes.}
\label{fig:grassmann_constraints}
\end{figure}

We find the final result for the full partition sum in the dual representation:
\begin{eqnarray}
Z & = &\frac{1}{2^{2V}} \!\!\sum_{\{p,k,\overline{k},s \}} \!\!\! C_{F}[s,k,\overline{k}\,]
\prod_{x,\mu} (-1)^{\, J_{x,\mu}^{\,21} + k_{x,\mu}^{\,21} + \overline{k}_{x,\mu}^{\,21}} \; 
\prod_{L} \, \mbox{sign} \,(L) \; \; W_M[s] \, W_\beta [p] \, 
W_H [p,k,\overline{k}]
\nonumber \\
& \times &  
\prod_{x, \mu} 
\delta \left( J_{x,\mu}^{\,11} \!+\! k_{x,\mu}^{\,11} \!-\! \overline{k}_{x,\mu}^{\,11} - 
\big[J_{x,\mu}^{\,22} \!+\! k_{x,\mu}^{\,22} \!-\! \overline{k}_{x,\mu}^{\,22} \big] \right)
\delta \left( J_{x,\mu}^{\,12} \!+\! k_{x,\mu}^{\,12} \!-\! \overline{k}_{x,\mu}^{\,12} - 
\big[J_{x,\mu}^{\,21} \!+\! k_{x,\mu}^{\,21} \!-\! \overline{k}_{x,\mu}^{\,21} \big] \right) 
\nonumber \\
& = & \frac{1}{2^{2V}} \!\!\sum_{\{p,k,\overline{k},s \}}  \;
\prod_{x,\mu} (-1)^{\, J_{x,\mu}^{\,21} + k_{x,\mu}^{\,21} + \overline{k}_{x,\mu}^{\,21}} \; 
\prod_{L} \, \mbox{sign} \,(L) \; (2m)^{\, \sum_{x,a} s_x^a} \;
\prod_{x,\mu<\nu} \prod_{a,b,c,d}  \dfrac{ \left( \frac{\beta}{2} \right)^{p_{x,\mu\nu}^{abcd}}}{p_{x,\mu\nu}^{abcd}!} 
\nonumber \\
& \times &
\prod_{x,\mu} \dfrac{\!\left(\!\frac{S_{x,\mu}^{11} +k_{x,\mu}^{\; 11}+\overline{k}_{x,\mu}^{\; 11}+ 
S_{x,\mu}^{22} +k_{x,\mu}^{\; 22}+\overline{k}_{x,\mu}^{\; 22}}{2}\!\right) ! \; 
 \left(\frac{S_{x,\mu}^{12} +k_{x,\mu}^{\; 12}+\overline{k}_{x,\mu}^{\; 12} + 
S_{x,\mu}^{21}+k_{x,\mu}^{\; 21}+\overline{k}_{x,\mu}^{\; 21}}{2}\right) !}{\!\left(\frac{
S_{x,\mu}^{11} + k_{x,\mu}^{\; 11}+\overline{k}_{x,\mu}^{\; 11} + 
S_{x,\mu}^{22} + k_{x,\mu}^{\; 22}+\overline{k}_{x,\mu}^{\; 22} + 
S_{x,\mu}^{12} + k_{x,\mu}^{\; 12}+\overline{k}_{x,\mu}^{\; 12} + 
S_{x,\mu}^{21} + k_{x,\mu}^{\; 21}+\overline{k}_{x,\mu}^{\; 21}  }{2} + 1\!\right) !} 
\nonumber \\
& \times & \prod_{x, \mu} 
\delta \left( J_{x,\mu}^{\,11} \!+\! k_{x,\mu}^{\,11} \!-\! \overline{k}_{x,\mu}^{\,11} - 
\big[ J_{x,\mu}^{\,22} \!+\! k_{x,\mu}^{\,22} \!-\! \overline{k}_{x,\mu}^{\,22} \big] \right)
\delta \left( J_{x,\mu}^{\,12} \!+\! k_{x,\mu}^{\,12} \!-\! \overline{k}_{x,\mu}^{\,12} - 
\big[ J_{x,\mu}^{\,21} \!+\! k_{x,\mu}^{\,21} \!-\! \overline{k}_{x,\mu}^{\,21} \big] \right)
\nonumber \\
& \times & 
\prod_{x} \prod_{a} \delta \left( 1 - s_{x}^{a} - \frac{1}{2} \sum_{\mu,b} \big[
k_{x,\mu}^{a b}  + k_{x-\hat{\mu},\mu}^{b a} + \overline{k}_{x-\hat{\mu},\mu}^{b a} + \overline{k}_{x,\mu}^{ab} \big] \right) 
.
\label{dualZfinal}
\end{eqnarray}
The partition sum is a sum over configurations of the dual variables for the fermions
$s_x^{a}, k_{x,\mu}^{\; ab}, \overline{k}_{x,\mu}^{\; ab} \in \{0,1\}$ and the cycle occupation numbers
$p_{x,\mu\nu}^{abcd} \in \mathbb{N}_{0}$ which represent the gauge field degrees of freedom. The constraint 
$C_{F}[s,k,\overline{k}\,]$ enforces admissible configurations, such that monomers, dimers and loops
completely fill the double layer lattice (in the last line of (\ref{dualZfinal}) we give an expression
for this constraint as a product of Kronecker deltas). The monomers come with the weight factor
$W_M[s]$, i.e., every monomer contributes a factor of $2m$ to the weight. The cycle occupation numbers 
give rise to weight factors $W_\beta [p]$ specified in (\ref{eq:weightfactorbeta}) and from the 
Haar measure integrals we find weight factors $W_H [p,k,\overline{k}]$ given in
(\ref{weighthaarfinal}). On all links of the lattice we have 
flux constraints where the corresponding Kronecker deltas connect the $11$ with the $22$ flux components, as well as 
the $12$ with the $21$ components. Note that in the presence of fermions this is now the combined flux 
from the ACCs which enter through the currents $J_{x,\mu}^{\,ab}$ and the fermionic fluxes coupling
via  $k_{x,\mu}^{\,ab} - \overline{k}_{x,\mu}^{\,ab}$ (see Fig.~\ref{fig:grassmann_constraints} for a graphical 
representation patterned after the pure gauge case shown in Fig.~\ref{fig:fluxconservation}.). 
Finally each configuration comes with signs which 
receive contributions from all fermion loops $L$ via the sign function sign$(L)$ given by (\ref{loopsign})
and sign factors counting the flux crossings. 

\section{Strong coupling limit, loop corrections and geometrical picture}

After having completed the derivation of the dual representation it is instructive to analyze the contributions to the
partition sum in its dual form. It is clear that such an analysis has to follow some organizational principle. As we will 
see in this section it is natural to start from the strong coupling limit, i.e., $\beta  = 0$ (compare \cite{Rossi:1984cv}), 
and then to consider finite $\beta$ corrections to that limit. 

\begin{figure}[t]
\begin{center}
\vskip5mm
\includegraphics[scale=0.6,clip]{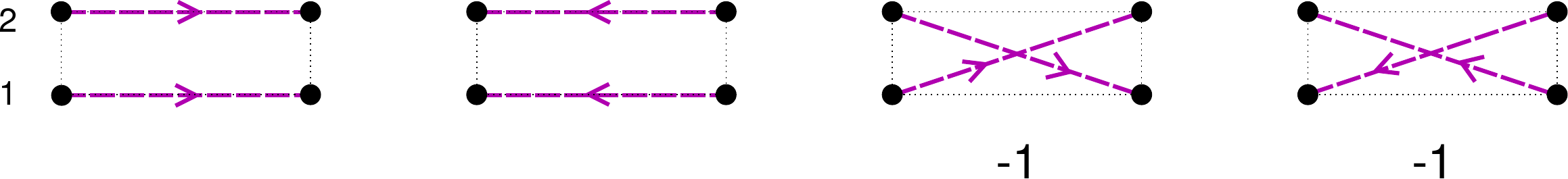}
\end{center}
\caption{Loop elements in the strong coupling limit. Only the four double fluxes shown here are admissible
for loops in the strong coupling limit. These strong coupling loops, together with monomers and the 
dimers shown in Fig.~\ref{fig:dimers} have to completely 
fill the lattice. The elements with crossing color flux come with an explicit minus sign.}
\label{fig:strong_loops}
\end{figure}

In the strong coupling limit all cycle occupation numbers  have to be zero, since from 
(\ref{eq:weightfactorbeta}) follows $\lim_{\beta \rightarrow 0} W_\beta[p] = 0$ unless $p_{x,\mu\nu}^{abcd} = 0 \; \, \forall
x,\mu,\nu,a,b,c,d$. Thus also the currents $J_{x,\mu}^{ab}$ all vanish and the constraints in (\ref{dualZfinal}) reduce to
\begin{equation}
k_{x,\mu}^{\,11} \!-\! \overline{k}_{x,\mu}^{\,11} \; = \; k_{x,\mu}^{\,22} \!-\! \overline{k}_{x,\mu}^{\,22}
\quad \forall x, \mu \qquad \mbox{and} \qquad 
k_{x,\mu}^{\,12} \!-\! \overline{k}_{x,\mu}^{\,12} \; = \; k_{x,\mu}^{\,21} \!-\! \overline{k}_{x,\mu}^{\,21}
\quad \forall x, \mu \; .
\label{strongconstraints}
\end{equation}
This implies that only special types of loops are admissible and can be used to fill the lattice together with monomers and 
dimers. Only four choices give rise to loop segments on the link $(x,\mu)$:  $k_{x,\mu}^{\,11}  =  k_{x,\mu}^{\,22}$,
$ \overline{k}_{x,\mu}^{\,11} = \overline{k}_{x,\mu}^{\,22}$, $k_{x,\mu}^{\,12} = k_{x,\mu}^{\,21}$,
$ \overline{k}_{x,\mu}^{\,12} = \overline{k}_{x,\mu}^{\,21}$. All four possibilities give rise to two units of 
flux in the same direction (forward or backward). The fluxes can run parallel in color space or cross, where the latter 
case gives rise to an explicit minus sign from the term $(-1)^{k_{x,\mu}^{21} + \overline{k}_{x,\mu}^{21}}$ for 
the flux crossings. These loop elements of the strong coupling limit are depicted in Fig.~\ref{fig:strong_loops}.

\begin{figure}[t]
\begin{center}
\vskip5mm
\includegraphics[scale=0.6,clip]{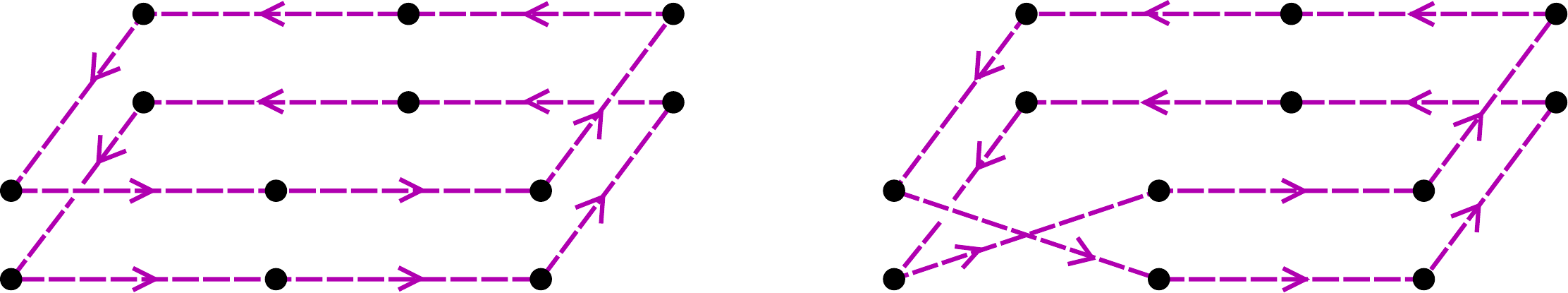}
\end{center}
\caption{Examples of fermion loops in the strong coupling limit. The double loop on the lhs.\ has a positive sign
since here the signs of the individual loops in the color 1 and the color 2 layers are the same and thus this sign 
squares to $+1$. The loop on the rhs.\ differs from the lhs.\ by one link element with a color crossing which comes with 
an explicit minus sign. On the other hand the color crossing also connects the two loops from the lhs.\ into a single 
loop such that there is one less minus sign from the number of loops. 
In total the loop on the rhs.\ thus also has a positive sign.}
\label{fig:example1}
\end{figure}

The configurations of the strong coupling limit thus consist of monomers, dimers as depicted in Fig.~\ref{fig:dimers}
and the strong coupling loops shown in Fig.~\ref{fig:strong_loops}. It is a remarkable property that all SU(2) 
strong coupling configurations are real and positive. This result is easy to see in our representation: First we 
remark again, that monomers and dimers all come with a positive sign and the only non-trivial signs come from the 
loops. Let us begin the analysis of the loop signs with considering a loop that is made from only the first two 
elements in Fig.~\ref{fig:strong_loops}, i.e., a double loop without color crossings 
(see the lhs.\ plot of Fig.~\ref{fig:example1} for an example). This situation corresponds to two loops running exactly 
parallel to each other in the two color layers. The signs sign($L$) of the two individual loops come only from the 
fermions and since these signs are independent of the color indices we have a positive overall sign 
sign($L)^2 = +1$ for strong coupling loops without color crossings. 

To study more general strong coupling loops we now replace one of the links by a strong coupling loop element 
with a crossing of color flux (the two elements shown on the rhs.\ of Fig.~\ref{fig:strong_loops}). An example of 
such a loop is shown on the rhs.\ of Fig.~\ref{fig:example1}. The flux crossing comes with an explicit minus sign. 
On the other hand, it also connects the two previously disconnected loops into a single loop, such that we have one
minus sign less (all other signs from the number of forward hops and from the staggered sign factors remain the same). 
Thus the overall sign of the new loop with a single flux crossing is again $+1$. In a similar way one can insert more additional
flux crossings. Each of them comes with an explicit minus sign but at the same time switches the connectivity between 
two disconnected loops and a single loop such that the overall sign always remains $+1$. In this way we can generate 
all possible strong coupling loops and conclude that all strong coupling configurations have a positive weight 
\cite{Rossi:1984cv}.

It is straightforward to include the leading corrections to the strong coupling limit. One starts to include loops where 
the gauge constraints are not saturated by fermion loop elements that run parallel, but instead by suitably activated ACCs. 
In order to illustrate such a calculation we display the leading terms of a coupled expansion of the partition sum 
in $1/m$ and in $\beta$:
\begin{eqnarray}
Z & = & m^{2V} \bigg[ 1 \; + \; \left( \!\frac{1}{2m}\! \right)^{\!2} \!\! \times \frac{1}{2!} \times 16 V \;+ \;   
\left( \! \frac{1}{2m}\! \right)^{\!4} \! \!\times \left( \!  \frac{1}{2!} \! \right)^{\! 2} \!\! \times [ 128 V^2 - 264 V] 
\; + \; \left(\! \frac{1}{2m}\! \right)^4 \!\!\times \frac{2!}{3!} \times 8 V 
\nonumber
\\
& + & 
\left(\! \frac{1}{2m}\! \right)^4 \!\!\times \frac{\beta}{2} \times  
\left( \!  \frac{1}{2!} \! \right)^{\! 4} \!\! \times 192 V \; + \; 
\left(\! \frac{\beta}{2}\! \right)^2 \!\! \times  
\left( \!  \frac{1}{2!} \! \right)^{\! 4} \!\! \times 48 V \; + \; 
{\cal O} \left(\!\left(\! \frac{1}{2m}\! \right)^{\!6\,}\!\right) +
{\cal O} \left(\beta^4\right) \bigg] .
\label{series}
\end{eqnarray}
The leading term in this expansion is the contribution where on all $2 V$
sites of the lattice the fermionic constraints are satisfied 
by placing $2 V$ monomers. The corresponding weight factor is $(2m)^{2V}$, which together with the
overall factor $(1/2)^{2V}$ gives rise to the contribution $m^{2V}$. We remark that on the rhs.\ of Eq.~(\ref{series})
we write the factor $m^{2V}$
up front, i.e., all further terms in the expansion are relative to the lattice completely filled with monomers. 

The next term is the configuration where we place a single dimer. For this contribution we need two monomers less,
such that we have a suppressing factor $(1/2m)^2$. Since here we have 
$k_{x,\mu}^{ab} = \overline{k}_{x,\mu}^{ab} = 1$ on a single link $(x,\mu)$ for some color combination $a,b$, 
we have a non-trivial factor $W_H[p,k,\overline{k}] = 1/2!$. Finally we have to compute the number of ways to place one 
dimer: On a given link we have 4 ways (compare Fig.~\ref{fig:dimers}) to place a dimer. Together with the number of links
$4 V$ this gives $16 V$ possibilities to place the dimer.

The next contribution is one where two dimers are placed on different links. It comes with a suppressing factor $(1/2m)^4$
and the $W_H[p,k,\overline{k}] = (1/2!)^2$, i.e., the squares of the corresponding factors for a single dimer. The number of
placements is the square of the degeneracy for a single dimer divided by two giving $(16 V)^2/2 = 128 V^2$. From this 
number we subtract the number of possibilities where two dimers touch (not admissible) or are placed at the same link. 

There is another possibility for placing two dimers: They can be on the same link, either parallel in color space, or crossing
diagonally.  Multiplied with the number of links this gives $2 \times 4 V = 8V$ possibilities. Here we have two $k$ and two 
$\overline{k}$ activated at the same link, such that we find $W_H[p,k,\overline{k}] = 2!/3!$. This contribution 
is the last term in the first   line of (\ref{series}).  

\begin{figure}[h]
\begin{center}
\vskip5mm
\includegraphics[scale=0.5,clip]{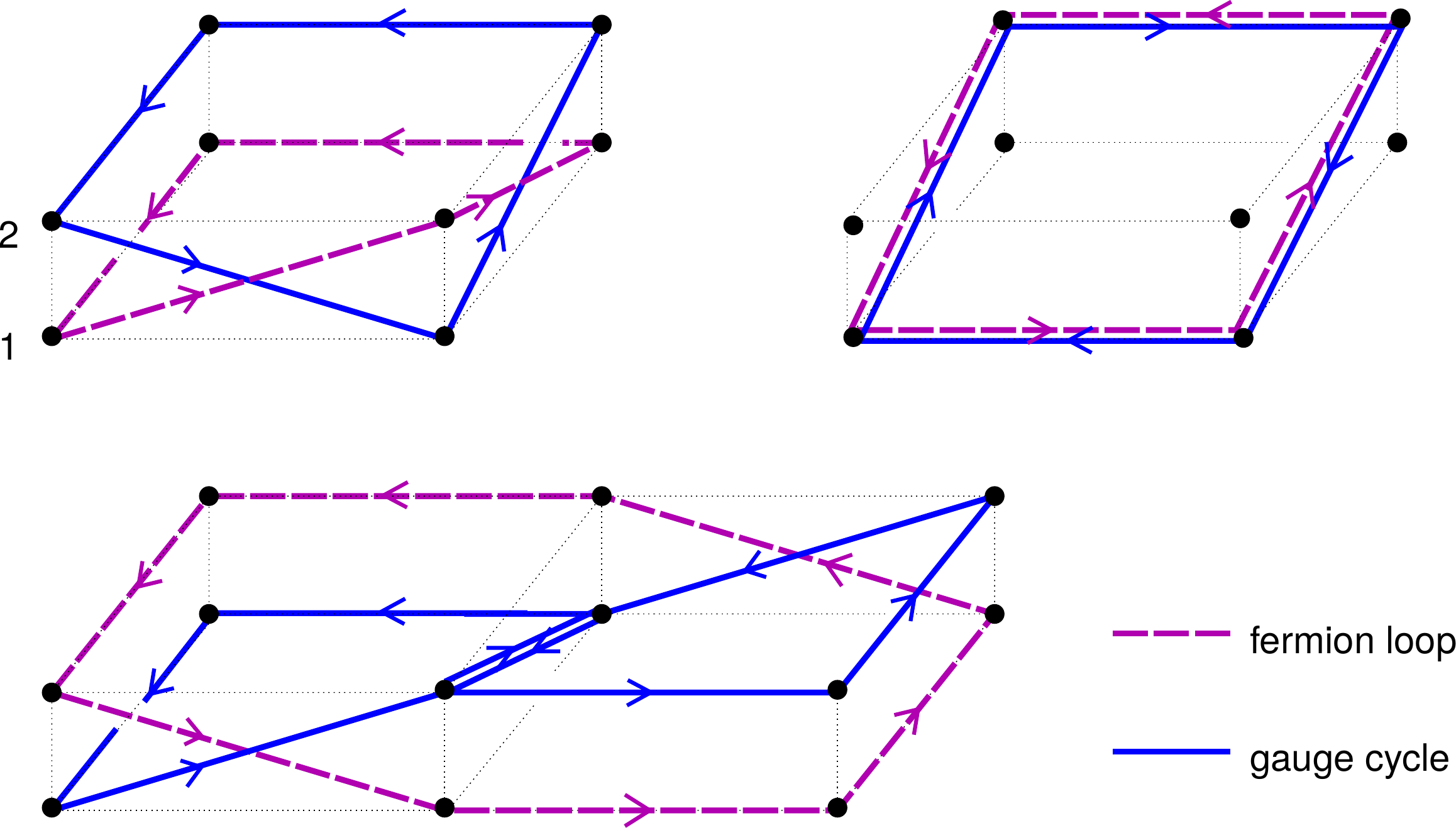}
\end{center}
\caption{Example for the saturation of a fermion loop with ACCs. \hfill}
\label{fig:example2}
\end{figure}

The first term in the second line of (\ref{series}) is the contribution of a loop around a plaquette, saturated with an ACC, thus 
coming with a factor of $\beta/2$ from the nontrivial cycle occupation number and a factor of $(1/2m)^4$ since four monomers
are replaced by the 4 corners of the loop around the plaquette. For the fermion loop around the plaquette we have 
two choices for the color the loop runs through at each corner giving a total of $2^4 = 16$ different loops. In addition each 
loop can have two orientations, and multiplying with $6 V$, the number of plaquettes, we find a degeneracy of 
$ 16 \times 2 \times 6 V = 192 V$. Note that the ACC which needs to be activated to compensate the flux from the fermion 
loop is uniquely determined. Since all ACCs are positively oriented, we have to distinguish two cases: 
For positively oriented loops (see the example on the lhs.\ in the top row of Fig.~\ref{fig:example2}) the ACC has the 
opposite color at each corner, while for negatively oriented fermion loops the ACC runs alongside the fermion
loop but with opposite orientation (rhs.~top of Fig.~\ref{fig:example2}). 
In both cases we have two units of flux on each link of the plaquette, such that we 
have the additional factor $W_H[p,k,\overline{k}] = (1/2!)^4$.

Finally the last term with $(\beta/2)^2$ 
in (\ref{series}) comes from placing two ACCs on top of each other such that the flux is compensated at 
each link. Since all ACCs are positively oriented, the two ACCs that match have opposite color at each 
corner, and we can form 8 different matching pairs from our 16 ACCs. Multiplied with 
$6 V$, the number of plaquettes, we find a degeneracy of $8 \times 6 V = 48 V$. Again we have two units of flux 
on each link of the plaquette, such that $W_H[p,k,\overline{k}] = (1/2!)^4$. Here all fermionic constraints are 
saturated with monomers.

We remark that some of the terms in (\ref{series}) were cross-checked with conventional techniques: The terms depending only
on $\beta$ match the corresponding terms of conventional strong coupling expansion \cite{Wilson:1974sk}, and for the $\beta$-independent terms the leading $(2m)^{-2}$ and $(2m)^{-4}$ contributions could be verified by comparison to the free case.
We furthermore stress at this point that the leading terms in the series (\ref{series}) are all positive and we found that negative signs
appear only at ${\cal O} \left(\beta^4\right)$ (see the example in Fig.~\ref{fig:unhappy}), or at   
${\cal O} \left(\!\left(\! \frac{1}{2m}\! \right)^{\!4\,} \! \beta^3\!\right)$ when a fermion loop is included.

The examples we have discussed for the leading terms of (\ref{series}) illustrate the use of the dual representation  
(\ref{dualZfinal}) for a combined strong coupling- and hopping expansion. The steps used for the leading terms can
easily be applied to more general contributions, and in the bottom of Fig.~\ref{fig:example2} we show in 
an example how to saturate the constraints for a larger fermion loop by filling it with ACCs. Actually this picture is
the same if one replaces the fermion loop by one of the closed color paths that contribute to the Wilson loop
($2 \times 1$ in this example) on the rhs.\ of (\ref{wilsonloop}). 

The discussion of the series expansion also illustrates the beautiful geometrical form of the dual representation 
(\ref{dualZfinal}). In the dual form the path integral is a theory of fluxes that have to obey the constraints on
all links as encoded in the Kronecker deltas of (\ref{dualZfinal}) and illustrated in Fig.~\ref{fig:grassmann_constraints}. 
The fluxes are either generated by fermion loops (more generally matter loops) or by the activation of abelian color cycles
which provide the fluxes from the gauge action around plaquettes. All sites that are not occupied by a fermion loop have
to be saturated with dimers or monomers. The latter come with factors $2m$, which together with the factors of 
$\beta/2$ for each unit of ACC flux and the combinatorial factors determine the weight of a configuration. 
The configurations come with a sign which is determined from the number of color crossings and the product of
loop signs sign$(L)$ given in (\ref{loopsign}).
 
\section{Discussion and outlook}

In this paper we have presented a new strategy for finding a dual representation for non-abelian lattice gauge theories, using
the gauge group SU(2) coupled to staggered fermions as our example. The approach is based on strong coupling expansion 
combined with a direct saturation of the Grassmann integral with monomers, dimers and loops. They key ingredient 
is a decomposition of the gauge action in terms of abelian color cycles (ACC) which are loops in color space
around plaquettes. The ACCs are abelian in nature, i.e., they commute, and the dualization can proceed as in the abelian 
case. The gauge fields can be integrated out completely and only simple combinatorial factors remain. 

The partition sum is exactly rewritten into a sum over loop, monomer and dimer configurations for the fermions and 
configurations over integer valued cycle occupation numbers for the gauge degrees of freedom. 
The loops and cycle occupation numbers are tied to each
other in constraints that live on the links of the lattice, and for each link require the cancellation of flux parallel in color 
space, as well as the cancellation of flux crossing in color space. From the gauge integral we obtain a minus sign for each
flux crossing and from the fermions the usual minus signs for staggered fermions. All other weight factors are real and 
positive and, as mentioned above, are simple combinatorial factors involving the cycle occupation numbers and loop and 
dimer occupation. Computing a few terms in a joint strong coupling and fermion loop/dimer expansion we find that the 
leading contributions all come with positive signs, i.e., fermionic signs and signs from color flux crossings cancel. However, 
at ${\cal O}(\beta^4)$ one finds negative sign contributions (at the moment we explore if these can be resummed). 
Observables have a simple representation in the dual language and we have discussed a few examples. 

The new approach based on ACCs opens up several interesting directions: It is obvious that the generalization to other 
gauge groups, in particular to SU(3), should be explored. At the moment we are studying the case of SU(3) and it is 
already clear that the ACC concept can be implemented as an extension of the SU(2) strategy developed here. 
Another direction 
which should be explored is the question whether a complete dualization is possible, in the sense that similar to the 
abelian case (see, e.g., \cite{Savit:1979ny}) another set of dual variables is introduced such that the constraints are 
automatically fulfilled. Finally it will be interesting to understand the deeper origin of the signs from the color flux crossing,
and in particular to explore the possibility of a resummation.

\vskip5mm
\noindent
{\bf Acknowledgements:}  We thank Falk Bruckmann for interesting discussions. This work is supported by the FWF DK W1203 
{\sl ''Hadrons in Vacuum, Nuclei and Stars''}, and partly also by the FWF Grant.\ Nr.\ I 1452-N27, as well as the
DFG TR55, {\sl ''Hadron Properties from Lattice QCD''}.

\end{document}